\documentclass[prd,showpacs,preprintnumbers,amsmath,amssymb,nofootinbib]{revtex4}

\usepackage{graphicx}
\usepackage{dcolumn}
\usepackage{bm}
\usepackage{dsfont}

\newcommand{\Od}{{\cal O}}

\newcommand{\tr}{\mbox{tr}}
\newcommand{\Tr}{\mbox{Tr}}

\newcommand{\diag}{\mbox{diag}}

\newcommand{\condtwo}{\langle \bar q q \rangle}
\newcommand{\condtwoi}{\langle \bar q_i q_i \rangle}
\newcommand{\condu}{\langle \bar u u \rangle}
\newcommand{\condd}{\langle \bar d d \rangle}
\newcommand{\conds}{\langle \bar s s \rangle}
\newcommand{\condsum}{\langle \bar u u + \bar d d\rangle}
\newcommand{\conddif}{\langle \bar u u - \bar d d\rangle}

\newcommand{\be}{\begin{equation}}
\newcommand{\ee}{\end{equation}}
\newcommand{\ba}{\begin{eqnarray}}
\newcommand{\ea}{\end{eqnarray}}

\newcommand{\gsim}{\raise.3ex\hbox{$>$\kern-.75em\lower1ex\hbox{$\sim$}}}
\newcommand{\lsim}{\raise.3ex\hbox{$<$\kern-.75em\lower1ex\hbox{$\sim$}}}

\begin{document}

\title{Isospin Breaking and chiral symmetry restoration}

\author{A. G\'omez Nicola}
\email{gomez@fis.ucm.es} \affiliation{Departamento de F\'{\i}sica
Te\'orica II. Univ. Complutense. 28040 Madrid. Spain.}
\author{R.Torres Andr\'es}
\email{rtandres@fis.ucm.es} \affiliation{Departamento de
F\'{\i}sica Te\'orica II. Univ. Complutense. 28040 Madrid. Spain.}

\begin{abstract}
We analyze  quark condensates and chiral (scalar) susceptibilities including
  isospin breaking effects at finite temperature $T$. These include $m_u\neq m_d$ contributions as well as
 electromagnetic ($e\neq 0$) corrections,  both treated in a consistent chiral lagrangian framework to leading order in
  $SU(2)$ and $SU(3)$ Chiral Perturbation Theory, so that our predictions are model independent.  The chiral restoration temperature
 extracted from $\condtwo=\condsum$ is almost unaffected, while the  isospin breaking
order parameter $\conddif$ grows with $T$ for the three-flavor case $SU(3)$. We derive a sum rule relating the condensate ratio $\condtwo (e\neq 0)/\condtwo(e=0)$ with the scalar susceptibility difference $\chi(T)-\chi(0)$, directly measurable on the lattice. This sum rule is useful also for estimating condensate errors in staggered lattice analysis. Keeping $m_u\neq m_d$ allows to obtain the connected and disconnected contributions to the susceptibility, even in the isospin limit, whose temperature, mass and isospin breaking  dependence we analyze in detail. The disconnected part grows linearly, diverging in the chiral (infrared) limit as  $T/M_\pi$, while the connected part shows a quadratic behaviour, infrared regular as $T^2/M_\eta^2$ and coming from $\pi^0\eta$ mixing terms. This smooth connected behaviour suggests that isospin breaking correlations are weaker than critical chiral ones near the transition temperature. We explore some consequences in connection with lattice data and their scaling properties, for which our present analysis for physical masses, i.e. beyond the chiral limit, provides a useful model-independent description for low and moderate temperatures.
   \end{abstract}



\maketitle

\section{Introduction}
The low-energy sector of QCD has been successfully described over recent years within the chiral lagrangian framework. Chiral Perturbation Theory (ChPT)
is based on the spontaneously breaking of chiral symmetry $SU_L(N_f)\times SU_R(N_f)\rightarrow SU_V(N_f)$ with $N_f=2,3$ light flavors and provides a consistent, systematic and model-independent scheme to calculate low-energy observables \cite{we79,Gasser:1983yg,Gasser:1984gg}.  The effective ChPT lagrangian is constructed as an expansion of the form ${\cal L}={\cal L}_{p^2}+{\cal L}_{p^4}+\dots$ where $p$ denotes a meson energy scale compared to the chiral scale $\Lambda_{\chi}\sim$ 1 GeV. For  $N_f=3$ case, the vector group symmetry is broken by the strange-light quark mass difference $m_s-m_{u,d}$, although $m_s$ can still be considered as a perturbation compared to  $\Lambda_{\chi}$, leading to $SU(3)$ ChPT, which reduces formally to $SU(2)$ in the $m_s\rightarrow\infty$ limit  \cite{Gasser:1984gg}. The formalism can also be extended to  finite temperature $T$, in order to describe
 meson gases and their evolution towards chiral symmetry restoration for $T$ below the critical temperature $T_c$ \cite{Gasser:1986vb,Gerber:1988tt}, where $T_c\simeq$ 180-200 MeV from lattice simulations \cite{Bernard:2004je,Aoki:2006we,Aoki:2009sc,Cheng:2009zi}. The use of ChPT in this context is important in order to provide model-independent results for the evolution of the different observables with $T$, supporting the original predictions for chiral restoration \cite{PisWil84}, also confirmed by lattice simulations, which are consistent with a crossover-like transition for $N_f=3$ (2+1 flavors in the physical case), which becomes  of second order  for $N_f=2$, in the $O(4)$ universality class, and first order in the degenerate case of three equal flavors.

   The $SU_V(2)$ vector group is the isospin symmetry, which is a very good approximation to Nature. However, there
 are several examples where isospin breaking corrections are phenomenologically relevant, such as sum rules for quark condensates \cite{Gasser:1984gg}, meson masses  \cite{Urech:1994hd} or  pion scattering \cite{Meissner:1997fa,Knecht:1997jw}.
 For a recent review see \cite{Rusetsky:2009ic}. The two possible sources of isospin breaking are  the QCD $m_d-m_u$ light quark mass difference and electromagnetic interactions. Both can be accommodated within the ChPT framework. The expected corrections from the first source are of  order $(m_d-m_u)/m_s$ and are encoded in the quark mass matrix,  generating also a  $\pi^0\eta$ mixing term in the $SU(3)$ lagrangian \cite{Gasser:1984gg}.  The  electromagnetic interactions are included in the ChPT effective lagrangian via the external source method and give rise to new terms \cite{Ecker:1988te,Urech:1994hd,Knecht:1997jw,Meissner:1997fa} of order ${\cal L}_{e^2}$, ${\cal L}_{e^2p^2}$ and so on,with $e$ the electric charge. These terms are easily incorporated in the ChPT power counting scheme by considering formally $e^2=\Od (p^2/F^2)$, with $F$ the pion decay constant in the chiral limit.

The purpose of this paper is to study within ChPT isospin
breaking effects  related to  the thermodynamics of the meson gas. We will be particularly interested in the physical quantities directly related to
spontaneous chiral symmetry breaking and its restoration, namely,
the quark condensates and their corresponding susceptibilities at finite temperature. The quark condensate is the order parameter of  chiral restoration, but since the transition is a smooth crossover for the physical case, different observables can yield different transition temperatures. Thus, the susceptibilities, defined as derivatives of the condensates with respect to the quark masses, provide also direct information about the transition and its nature, since they tend to peak around the transition point reflecting the growth of correlations.

Let us mention some of the motivations we have in mind for the present analysis. For the physical values of quark and meson masses, we are interested
 in the effect of the isospin-breaking terms in the light quark condensate $\condsum$ and therefore on the ChPT estimates of the critical
  temperature. In addition,  in the isospin asymmetric case, one has
$\condu\neq \condd$ and in fact $\conddif$ can be considered
an order parameter
  for isospin breaking. Actually, isospin is not spontaneously broken in QCD \cite{Vafa:1983tf} which means that this order parameter should vanish for $m_u=m_d$ and $e=0$. This is an important difference with the scalar condensate $\condsum$, which is nonzero in the chiral limit.
  It is  relevant to estimate the thermal evolution of
$\conddif$,  since
in principle the two condensates  melt at different critical
temperatures. A further motivation  is the analysis of  the three independent susceptibilities,
 directly related to the isosinglet, connected (isotriplet) and disconnected susceptibilities \cite{Smilga:1995qf} often discussed in  lattice analysis \cite{Detar:2007as,DeTar:2009ef,Ejiri:2009ac,Unger:2009zz}. Including properly the  $m_u-m_d$ dependence of condensates is then essential to analyze the temperature and mass evolution of the connected and disconnected pieces measured in the lattice. In particular, the linear
$m_d-m_u$ corrections to condensates survive the $m_u=m_d$ limit in the susceptibilities. The contributions coming from $\pi^0\eta$ mixing in
the $SU(3)$ case belong to this type and are particularly important regarding the temperature dependence. This is not only interesting for physical masses but also to explore the scaling near the chiral limit, which in lattice studies has been used to investigate the nature of the transition \cite{Ejiri:2009ac}. In the lattice works, this scaling may be contaminated by lattice artifacts such as taste breaking in the staggered fermion formalism, which can generate contributions to susceptibilities masking the true scaling behaviour \cite{Ejiri:2009ac,Unger:2009zz}. Our study provides then a model-independent setup for disentangling these effects and establishes the expected results in the continuum limit.

We will work in ChPT to one loop, considering on the same footing the two sources of
isospin-breaking. In a previous work \cite{Nicola:2010xt} we have studied the quark condensates at $T=0$
 and several related phenomenological aspects of the isospin asymmetric case. We will refer to that work for more details about the formalism, the numerical values of the low-energy constants and other related issues.

   The paper  is organized as follows. In section \ref{sec:form} we will review the main aspects of the isospin-breaking ChPT formalism related to the present work. Our results for the quark condensates at finite $T$ both in the $SU(2)$ and $SU(3)$ cases are given and analyzed in section \ref{sec:cond}. In that section we explore the temperature dependence of isospin breaking, as well as that of the sum rule relating condensate ratios. Section \ref{sec:susc} is devoted to the analysis of the different isospin-breaking scalar susceptibilities and their relation to the connected and disconnected ones. In subsection \ref{sec:sumrule} we provide an interesting sum rule relating the electromagnetic differences in the condensates with the total susceptibility. We explore the possibility of using that sum rule to estimate the errors in the staggered fermion lattice analysis of the condensates, in connection with the taste breaking effect. In subsection \ref{sec:sustempmass} we make a thorough study of the connected and disconnected contributions to the susceptibility and their dependence with temperature, the quark mass and the isospin ratio $m_u/m_d$. We pay special attention to the connection of our results with different lattice analysis in the literature.

\section{Formalism}
\label{sec:form}

The effective chiral lagrangian up to fourth order in $p$ (a meson mass, momentum, temperature or derivative) including electromagnetic interactions proportional to $e^2$ is given schematically by ${\cal L}_{eff}={\cal L}_{p^2+e^2}+{\cal L}_{p^4+e^2p^2+e^4}$.  The most general second order lagrangian is the familiar non-linear sigma model, including  the gauge coupling of mesons to the electromagnetic field through the covariant derivative, plus an additional  term proportional to a low-energy constant $C$ compatible with the $e\neq 0$ symmetries of the QCD lagrangian \cite{Ecker:1988te,Urech:1994hd}:

\begin{equation}
{\cal L}_{p^2+e^2}=\frac{F^2}{4} \tr\left[D_\mu U^\dagger D^\mu U+2B_0{\cal M}\left(U+U^\dagger\right)\right]+C\tr\left[QUQU^\dagger\right].
\label{L2}
\end{equation}

Here, $U(x)=\exp [i\Phi/F]\in SU(N_f)$, with $\Phi$  the Goldstone Boson (GB) matrix field for pions ($N_f=2$) plus kaons and $\eta$ ($N_f=3$), the latter being the octet member with $I_3=S=0$. The covariant
derivative is $D_\mu=\partial_\mu+iA_\mu[Q,\cdot]$ with $A_\mu$ the EM
field. ${\cal M}$ and $Q$ are the quark mass and charge matrices,
i.e., in $SU(3)$ ${\cal M}=\diag (m_u,m_d,m_s)$ and
$Q=(e/3)\diag(2,-1,-1)$. Both the mass term and the charge one proportional to $C$ in (\ref{L2}) break explicitly the chiral symmetry $SU_L(N_f)\times SU_R(N_f)$ under which $U\rightarrow LUR^\dagger$ with $L,R\in SU(N_f)$. The vector symmetry $L=R$ is also broken for unequal quark masses and charges. Thus, in the light sector ($u,d$) the part of the mass term proportional to $\hat m=(m_u+m_d)/2$, the average light quark mass, is also proportional to the identity flavor matrix and therefore invariant under $SU_V(2)$, while the part proportional to the mass difference $m_\delta=(m_u-m_d)/2$ and $T_3$, the third isospin generator, is the one carrying out the QCD isospin breaking.   The only remaining symmetry of the lagrangian (\ref{L2}) is the $U(1)$ $L=R=\exp(i\lambda Q)$ corresponding to charge conservation.

Working out the kinetic terms in (\ref{L2}) allows to relate the low-energy parameters  $F,B_0 m_{u,d,s}, C$ to the leading-order tree level values for the  decay constants  and masses of the pseudo-Goldstone bosons. For $SU(2)$ the masses read:

\begin{eqnarray}
M_{\pi^+}^2&=&M_{\pi^-}^2=2\hat m B_0 + 2 C \frac{e^2}{F^2},\nonumber\\
M_{\pi^0}^2&=& 2\hat m B_0 \label{treemassessu2}.
\end{eqnarray}

In the $SU(3)$ case,  the mass term in (\ref{L2}) induces a mixing  between the $\pi^0$ and the $\eta$ fields given by ${\cal L}_{mix}=(B_0/\sqrt{3})(m_d-m_u)\pi^0\eta$. This mixing between the two states with $I_3=S=0$ will play an important role in what follows.
The kinetic term has then to be brought to the canonical form before identifying the GB masses, which can be easily done by the field rotation \cite{Gasser:1984gg}:

\begin{eqnarray}
\pi^0&=&\bar\pi^0\cos\varepsilon-\bar\eta\sin\varepsilon,\nonumber\\
\eta&=&\bar\pi^0\sin\varepsilon+\bar\eta\cos\varepsilon,
\label{pietarot}
\end{eqnarray}
where the mixing angle is given by:

\begin{equation}
\tan 2\varepsilon=\frac{\sqrt{3}}{2}\frac{m_d-m_u}{m_s-\hat m}.
\label{mixangle}
\end{equation}

Once the above $\pi^0\eta$ rotation is performed, the $SU(3)$ tree level meson masses to leading order read:

\begin{eqnarray}
M_{\pi^+}^2&=&M_{\pi^-}^2=2\hat m  B_0 + 2 C \frac{e^2}{F^2},\nonumber\\
M_{\pi^0}^2&=&  2B_0\left[\hat m -\frac{2}{3} (m_s-\hat m)\frac{\sin^2\varepsilon}{\cos 2\varepsilon}\right],\nonumber\\
M_{K^+}^2&=&M_{K-}^2=(m_s+m_u) B_0 + 2 C \frac{e^2}{F^2},\nonumber\\
M_{K^0}^2&=&(m_s+m_d) B_0,\nonumber\\
M_\eta^2&=&2B_0\left[\frac{1}{3}(\hat m + 2m_s)+\frac{2}{3}(m_s-\hat m)\frac{\sin^2\varepsilon}{\cos 2\varepsilon}\right].
\label{treemassessu3}
\end{eqnarray}

For pions, the main effect in the $\pi^0-\pi^+$ mass difference comes from the EM contribution \cite{Das:1967it}, while in the kaon and eta cases the violations of Dashen's theorem $M_{K^\pm}^2-M_{K^0}^2=M_{\pi^\pm}^2-M_{\pi^0}^2$ \cite{Dashen:1969eg} ($m_u=m_d$ limit) indicate that $m_u-m_d$ corrections are relevant and must be kept on the same footing as the EM ones \cite{Urech:1994hd,Bijnens:1996kk}. We emphasize that all the previous expressions hold for tree level LO masses $M_a^2$ with $a=\pi^{\pm},\pi^0,K^{\pm},\eta$, in terms of which we will  express all our results.  They
 coincide with the physical masses
 to leading order in ChPT, i.e., $M_{a,phys}^2=M_a^2(1+\Od(M^2))$ and so on for the meson decay constants $F_a^2=F^2(1+\Od(M^2))$.

The fourth-order lagrangian  consists of all possible terms compatible with the QCD symmetries to that order, including the EM ones. The ${\cal L}_{p^4}$ lagrangian is given in \cite{Gasser:1983yg} for the $SU(2)$ case, $h_{1,2,3}$ (contact terms) and $l_{1\dots 7}$ denoting the dimensionless low-energy constants (LEC) multiplying each independent term, and in \cite{Gasser:1984gg} for $SU(3)$  the LEC named  $H_{1,2}$ and $L_{1\dots 10}$. The electromagnetic ${\cal L}_{e^2p^2}$ and ${\cal L}_{e^4}$ for $SU(2)$ are given in \cite{Meissner:1997fa,Knecht:1997jw}, $k_{1,\dots 13}$ denoting the corresponding LEC, and in \cite{Urech:1994hd} for $SU(3)$ with the $K_{1\dots17}$ LEC. The relevant terms needed for this work are given in \cite{Nicola:2010xt}.

The LEC are renormalized in such a way that they absorb all the one-loop ultraviolet divergences coming from ${\cal L}_{p^2}$ and ${\cal L}_{e^2}$, according to the ChPT counting, rendering the observables   finite and scale-independent. The numerical values of the LEC at a given scale can be fitted to meson experimental data, except the contact  $h_i$ and $H_i$. The latter are needed for renormalization but cannot be directly measured,  reflecting an ambiguity in the  observables depending on them. The origin of this ambiguity is in the very same definition of the condensates in perturbation theory \cite{Gasser:1983yg}. It is therefore convenient to define suitable combinations which are independent of those constants and therefore can be determined numerically. We will bear this in mind throughout this work and we will try to provide such combinations when isospin-breaking is included. The numerical values we will use for masses and low-energy constants in the $SU(3)$ case are the same as in \cite{Nicola:2010xt} unless otherwise stated. In  $SU(3)$ they come from the fits performed in \cite{Amoros:2001cp}.

\section{Quark condensates at finite temperature}
\label{sec:cond}

The quark condensates   for a given flavor $q_i$ at finite temperature $T$ are given by:

\begin{eqnarray}
 \condtwoi_T= -\frac{1}{\beta V}\frac{\partial}{\partial m_i}\log Z=\left\langle \frac{\partial {\cal L}_{eff}}{\partial m_i}\right\rangle_T ,
\label{conddef}
\end{eqnarray}
where $\beta=1/T$, $V$ is the system volume, $Z$ the partition function and  $\langle\cdot\rangle_T$ denotes a thermal average. We will denote by $\condtwo_T=\condsum_T=-\frac{1}{\beta V}\frac{\partial}{\partial \hat m}\log Z$, the order parameter of chiral symmetry, while $\conddif_T=-\frac{1}{\beta V}\frac{\partial}{\partial m_\delta}\log Z$ behaves as an order parameter of isospin breaking, since it is the expectation value of the part of the mass term in the QCD lagrangian proportional to $m_u-m_d$ and $e(q_u-q_d)$, respectively. It is still invariant under transformations in the  third direction of isospin, which reflects electric charge conservation.

In ChPT to one loop we obtain then the  $SU(2)$ finite temperature extension of the $T=0$ results in \cite{Nicola:2010xt}, which we give also for consistency:

  \begin{eqnarray}
\condtwo_T&\equiv& \langle \bar u u + \bar d d \rangle_T=\condtwo_0+B_0\left[g_1(M_{\pi^0},T)+2g_1(M_{\pi^\pm},T)\right]+\mathcal{O}\left(p^2\right),\nonumber\\
\condtwo_0&=&-2F^2B_0\left[1-\mu_{\pi^0}-2\mu_{\pi^\pm}+2\frac{M_{\pi^0}^2}{F^2}\left(l_3^r(\mu)+h_1^r(\mu)\right) +e^2{\cal K}_2^r(\mu)+\mathcal{O}\left(p^4\right)\right],
\label{condsu2sum}
  \end{eqnarray}

\begin{equation}
\langle \bar u u - \bar d d \rangle_T=\langle \bar u u - \bar d d \rangle_0=4B_0^2(m_d-m_u)h_3-\frac{8}{3}F^2B_0e^2k_7 + \mathcal{O}\left(p^2\right),
\label{condsu2dif}
  \end{equation}
where:
\begin{equation}
{\cal K}_2^r(\mu)=\frac{4}{9}\left[5\left(k_5^r(\mu)+k_6^r(\mu)\right)+k_7\right],
\end{equation}

and

 \begin{eqnarray}
  \mu_i&=&\frac{M_i^2}{32\pi^2 F^2}\log\frac{M_i^2}{\mu^2},\nonumber\\
 g_1(M,T)&=&\frac{1}{2\pi^2}\int_0^\infty dp \frac{p^2}{E_p} \frac{1}{e^{\beta E_p}-1},
 \label{defsmug1}
\end{eqnarray}
with $E_p^2=p^2+M^2$.

The expression (\ref{condsu2sum})  contains the leading order tree level term from ${\cal L}_2$ given by $\condtwo_0=-2F^2B_0$, the one-loop tadpole like contribution $G_i(x=0)$, with $G$ the free meson thermal propagator, whose finite part yields the combinations $\mu_i+g_1(M_i,t)/(2F^2)$ (we follow the same finite-$T$ notation as in \cite{Gerber:1988tt}) and the tree level from the fourth order lagrangian, showing up only at $T=0$, which contains the LEC renormalized at the
 scale $\mu$ of dimensional regularization in the $\overline{MS}$ scheme \cite{Gasser:1983yg,Knecht:1997jw} so that the full expressions for the condensates are finite and scale-independent. Note that $\condtwo_0$ includes the contact term $h_1^r$.  The isospin breaking in $\condtwo_T$ for $SU(2)$ is purely electromagnetic, showing up explicitly in the $e^2$ terms and implicitly through the pion mass differences. The temperature dependence is encoded in the functions $g_1(M,T)$ which increase with $T$ and behave near the chiral limit ($T\gg M$) as $g_1(M,T)=\frac{T^2}{12}[1+\Od(M/T)]$.

 Note that the effect of the electromagnetic corrections is to decrease the thermal part of  $\condtwo_T$, since $M_{\pi^\pm}>M_{\pi^0}$. On the other hand, $\condtwo_0$  increases for the available estimates of the EM LEC, reflecting its ferromagnetic nature \cite{Nicola:2010xt}. Our first  conclusion is then that the critical temperature, estimated as that for which the condensate vanishes, increases with respect to the $e=0$ case, which is also a ferromagnetic-like behaviour induced by the explicit chiral symmetry breaking of the EM quark coupling in the QCD action. A simple estimate of the size of this effect can be obtained by taking the chiral limit $m_u=m_d=0$ so that $M_{\pi^0}=\mu_{\pi^0}=0$, $M_{\pi^\pm}^2=2Ce^2/F^2$ and   $T_c=\sqrt{8F}\sqrt{1+e^2{\cal K}_2^r-2\mu_{\pi^\pm}}$, which gives $T_c^{e\neq0}/T_c^{e=0}\simeq 1.003$ with the parameters used in \cite{Nicola:2010xt} and setting the involved $k_i$ to their maximum expected ``natural" values $k_i=1/(16\pi^2)$. Thus, in principle we expect rather small corrections to chiral restoration from the electromagnetic breaking. Nevertheless, in section \ref{sec:sumrule} we will go back to this point in connection with a sum rule relating the charge breaking with the susceptibility, suggesting larger corrections either for higher order transitions or for finite lattice spacing.

The two sources of explicit isospin breaking in the lagrangian show up in the condensate difference (\ref{condsu2dif}), which depends linearly on $m_u-m_d$ with the contact $h_3$ and vanishes for $m_u=m_d$ and $e=0$ in accordance with the absence of  spontaneous isospin breaking \cite{Vafa:1983tf} mentioned in the introduction. Recall that $h_3$ and $k_7$ do not need to  be  renormalized and are therefore finite and scale-independent. An important point is that $\conddif$ does not receive pion loop corrections in the two-flavor case and it is therefore temperature independent to the one-loop order. In other words, isospin breaking in $SU(2)$ does not change with $T$ and the two condensates melt at the same temperature. This picture will change for $N_f=3$ due to kaon loops and  $\pi^0\eta$ mixing.

In the $SU(3)$ case, we  calculate to one loop at finite temperature the light and strange condensates, taking into account both $m_u-m_d$ and $e\neq 0$ corrections.  The condensates read now:

\begin{eqnarray}
  \condtwo_T^{SU(3)}\equiv \langle \bar u u + \bar d d \rangle_T^{SU(3)}&=&\condtwo_0^{SU(3)} +B_0\left[\frac{1}{3}\left(3-\sin^2 \varepsilon\right) g_1(M_{\pi^0},T)+2g_1(M_{\pi^\pm},T)+g_1(M_{K^0},T)+g_1(M_{K^\pm},T)\right.\nonumber\\&+&\left.\frac{1}{3}\left(1+\sin^2 \varepsilon\right)g_1(M_\eta,T)\right]+\mathcal{O}\left(p^2\right),\nonumber\\
  \condtwo_0^{SU(3)} &=&-2F^2B_0\left\{1+\frac{8B_0}{F^2}\left[\hat m\left(2L_8^r(\mu)+H_2^r(\mu)\right)+4(2\hat m + m_s)L_6^r(\mu)\right]+e^2\mathcal{K}_{3+}^r (\mu)\right.\nonumber\\&-&\left.\frac{1}{3}\left(3-\sin^2 \varepsilon\right) \mu_{\pi^0}-2\mu_{\pi^\pm}-\mu_{K^0}-\mu_{K^\pm}-\frac{1}{3}\left(1+\sin^2 \varepsilon\right)\mu_\eta +\mathcal{O}\left(p^4\right)\right\},
\label{condsu3sum}
\end{eqnarray}

\begin{eqnarray}
\langle \bar u u - \bar d d \rangle_T^{SU(3)}&=&\langle \bar u u - \bar d d \rangle_0^{SU(3)}+B_0\left\{\frac{\sin 2\varepsilon}{\sqrt{3}}\left[g_1(M_{\pi^0},T)-g_1(M_\eta,T)\right]+g_1(M_{K^\pm},T)-g_1(M_{K^0},T)\right\}+\mathcal{O}\left(p^2\right),
\nonumber\\
\langle \bar u u - \bar d d \rangle_0^{SU(3)}&=&2F^2B_0\left\{\frac{4B_0}{F^2}(m_d-m_u)\left(2L_8^r(\mu)+H_2^r(\mu)\right)-e^2\mathcal{K}_{3-}^r (\mu)
\right.\nonumber\\&+&\left.
 \frac{\sin 2\varepsilon}{\sqrt{3}}\left[\mu_{\pi^0}-\mu_{\eta}\right]+\mu_{K^\pm}-\mu_{K^0}\right\}+\mathcal{O}\left(p^2\right),
\label{condsu3dif}
\end{eqnarray}

\begin{eqnarray}
\langle \bar s s \rangle_T&=&\langle \bar s s \rangle_0+B_0\left\{\frac{2}{3}\left[g_1(M_{\pi^0},T)\sin^2 \varepsilon+g_1(M_\eta,T)\cos^2 \varepsilon\right]+g_1(M_{K^\pm},T)+g_1(M_{K^0},T)\right\}+\mathcal{O}\left(p^2\right),\nonumber\\
\langle \bar s s \rangle_0&=&-F^2B_0\left\{1+\frac{8B_0}{F^2}\left[m_s\left(2L_8^r(\mu)+H_2^r(\mu)\right)+4(2\hat m + m_s)L_6^r(\mu)\right]+e^2\mathcal{K}_{s}^r (\mu)
\right.\nonumber\\&-&\left.
\frac{4}{3}\left[ \mu_{\pi^0}\sin^2 \varepsilon+\mu_{\eta}\cos^2 \varepsilon\right]-2\left[\mu_{K^\pm}+\mu_{K^0}\right]+\mathcal{O}\left(p^4\right)\right\},
  \label{condsu3str}\end{eqnarray}
where:

\begin{eqnarray}
\mathcal{K}_{3+}^r(\mu)&=&\frac{4}{9}\left[6\left(K_7+K_8^r(\mu)\right)+5\left(K_9^r(\mu)+K_{10}^r(\mu)\right)\right],\nonumber\\
\mathcal{K}_{3-}^r(\mu)&=&\frac{4}{3}\left[K_9^r(\mu)+K_{10}^r(\mu)\right],\nonumber\\
\mathcal{K}_{s}^r(\mu)&=&\frac{8}{9}\left[3\left(K_7+K_8^r(\mu)\right)+K_9^r(\mu)+K_{10}^r(\mu)\right].
\label{Kscondsu3}
\end{eqnarray}

In some of the above terms we have preferred to leave the results in terms of quark  instead of meson masses. As in the $SU(2)$ case, the results are finite and scale-independent, which concerns only the $T=0$ part \cite{Nicola:2010xt}.

 There are some important differences with respect to the $N_f=2$ case which deserve to be commented. First, the presence of the $\pi^0\eta$ mixing angle $\varepsilon$ (\ref{mixangle}), as well as the more complicated dependence of meson masses with quark masses (\ref{treemassessu3}), imply that now  $m_u-m_d$ corrections show up in $\condtwo$, apart from the EM ones. Note also that these corrections in $\condtwo$ and $\conds$ are at least $\Od(\varepsilon^2)$ in the mixing angle, or equivalently in $m_u-m_d$, except for an $\Od(e^2\varepsilon)$ term in the kaon contribution. This is so because, apart from the explicit $\varepsilon$ dependence, one has to expand also the meson masses in (\ref{treemassessu3}) around $\varepsilon=0$. All the masses depend quadratically on $\varepsilon$ except $M_{K^\pm}^2\sim-a\varepsilon$, $M_{K^0}^2\sim a\varepsilon$ with $a=(2B_0/\sqrt{3})(m_s-\hat m)$. Since, in addition,  $M_{K^\pm}^2=M_{K^0}^2+2Ce^2/F^2$ for $\varepsilon=0$, we end up with the above mentioned term.

 Another important difference between the two cases is that for $SU(3)$ there are loop contributions to $\conddif_T$ in (\ref{condsu3dif}). Kaon loops arise from the charged-neutral kaon mass difference, while neutral pion and eta ones from $\pi^0\eta$ mixing. When expanding in $\varepsilon$ now, the leading order is $\Od(\varepsilon)$ even for $e=0$. These linear terms will be crucial for our analysis of susceptibilities in section \ref{sec:susc}. Those loop corrections introduce now a $T$ dependence in $\conddif_T$, unlike the $SU(2)$ case. As it happened in the $SU(2)$ case, we see  that $\conddif_T$ in (\ref{condsu3dif}) vanishes for $e^2$ and $m_u=m_d$, in agreement with  \cite{Vafa:1983tf}, which we see from our analysis that holds including  thermal corrections.

At low and moderate temperatures $g_1(M_{\pi^0},T)$ dominates over the kaon and eta contributions in (\ref{condsu3dif}), but it should be reminded that $\varepsilon$ in (\ref{mixangle}) brings up a $1/m_s$ dependence which reduces the size of the pion term. In order to make a crude estimate, let us consider again the chiral limit, but keeping now the leading order in $m_u-m_d$, which we take then very small but nonzero while taking $\hat m\rightarrow 0^+$. In this limit the kaon masses are roughly kept to their physical values, which are well above the critical temperature.  Thus, we consider the regime $M_\pi\ll T \ll M_K$, in which the pion term behaves as $B_0(m_d-m_u)T^2/(24m_s)=B_0^2(m_d-m_u)T^2/(18M_\eta^2)$. The kaon and eta  contributions  go like  $B_0T^2\left[(m_d-m_u)/m_s\right]\sqrt{M_{K,\eta}/T}e^{-M_{K,\eta}/T}$  \cite{Gerber:1988tt}, where we have taken also $e=0$ for simplicity. The pion term is still dominant due to the exponential suppression of $K$,$\eta$. However, when compared to the $T=0$ part in that regime, which goes like $(m_d-m_u)B_0^2$, we see that the quadratic growth with temperature is controlled by the scale $M_\eta^2$ instead of, say, the chiral restoring behaviour of $\condtwo_T$ which is controlled by $F^2$ in the chiral limit. Therefore, the order parameter for isospin breaking $\conddif_T$ grows with $T$, although it does so rather softly. Therefore, we do not expect big differences in the melting temperatures of the $u$ and $d$ condensates. This is also consistent with the expectation that in the limit where $m_s$ is arbitrarily large, say compared to $\hat m$, the $SU(2)$ result should be recovered, for which there is no temperature dependence for the condensate difference.

\begin{figure}[h]
\includegraphics[scale=.41]{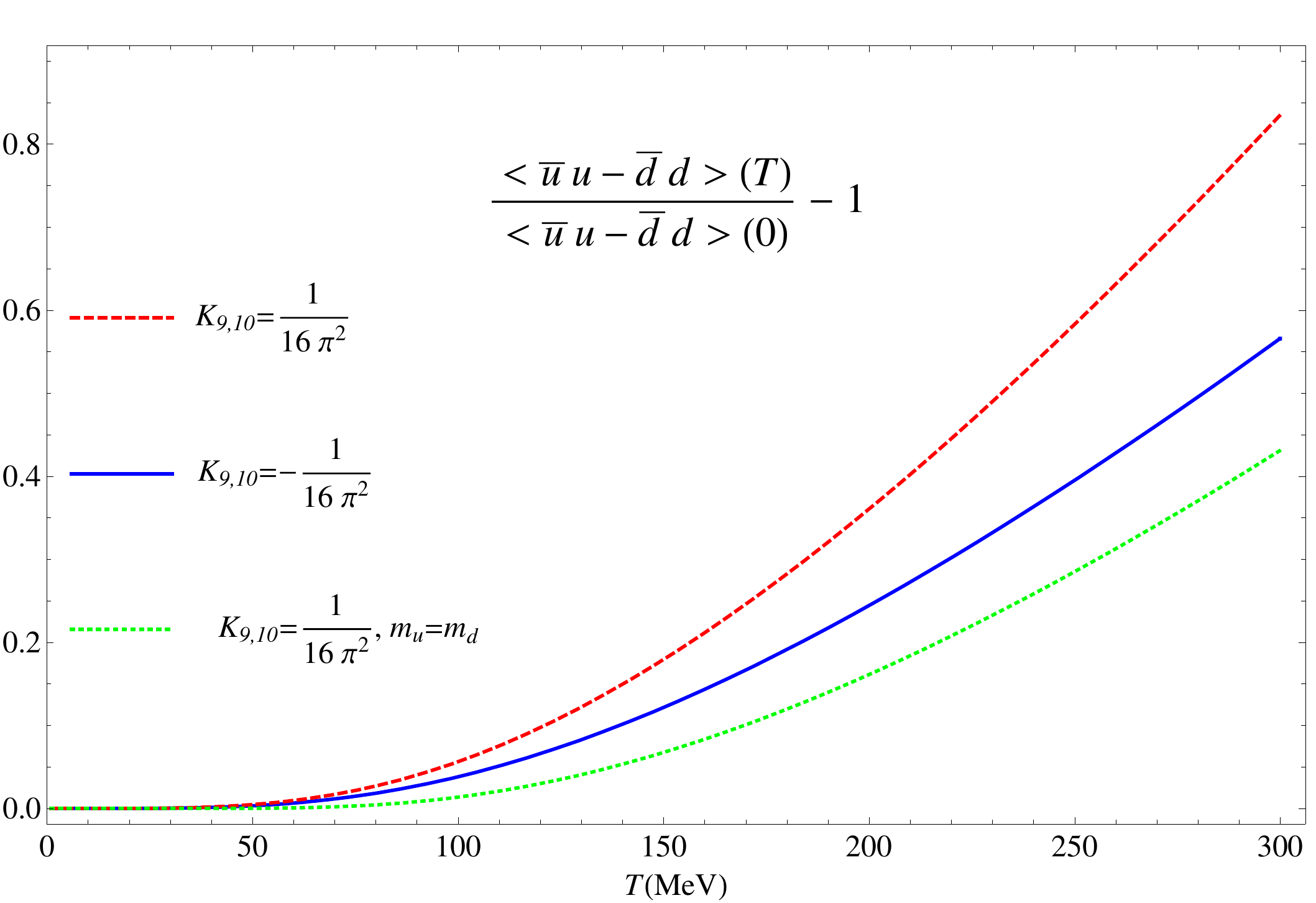}
\includegraphics[scale=.5]{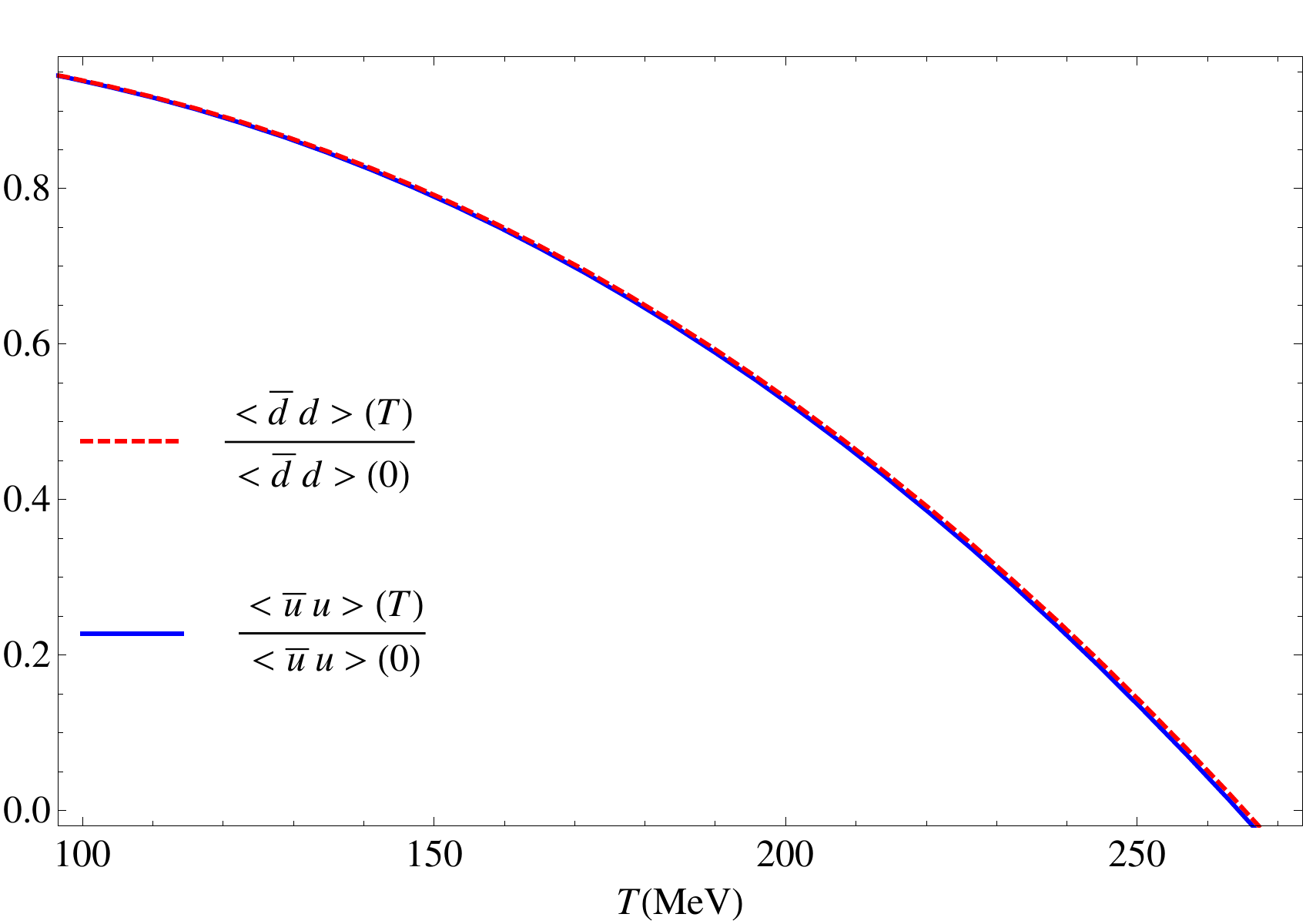}
 \caption{\rm \label{fig:uddifsu3} Left: The $u-d$ condensate difference (isospin breaking order parameter) at finite temperature in $SU(3)$, relative to its $T=0$ value. Right: The two condensates separately.}
\end{figure}

The evolution
with  temperature of the condensate difference is shown in Figure \ref{fig:uddifsu3} for the full case of finite
 pion mass and both $e\neq 0$ and $m_u\neq m_d$. We have used the same set of low-energy constants and parameters
  as in \cite{Nicola:2010xt}, in particular $m_u/m_d=0.46$ and $m_s/\hat m=24$. For the EM LEC $K_i$ involved, we have displayed in the figure
   the two curves corresponding to their maximum and minimum expected natural values. We also show for comparison the result for $m_u=m_d$, which
    shows that the charge contribution is actually of the same order as the one proportional to $m_u-m_d$. We see that the $T$-dependent  amplification of  the isospin difference is rather large. In fact, this order parameter reaches values comparable to its $T=0$ value near the critical temperature, which is about $T_c\simeq$ 265 MeV
in $SU(3)$ ChPT. Nevertheless, due to the additional $\varepsilon$ suppressing factor discussed above, this enhancement is not enough to
produce a sizable difference in the melting temperature of the
$u,d$ condensates, as it is clearly seen in Figure
\ref{fig:uddifsu3} (right), where we plot the two thermal
condensates separately. The two plots showed in Figure \ref{fig:uddifsu3} correspond then respectively to the two order parameters involved here: isospin breaking and chiral restoration. In turn, note that the    curves on the right plot are independent of the
choice of LEC since to this order
$\condtwoi_T/\condtwoi_0=1-(\condtwoi_T-\condtwoi_0)/(B_0
F^2)+\Od(p^4)$ for $i=u,d$.

The individual condensates in (\ref{condsu3sum})-(\ref{condsu3str})  contain the contact terms $H_2$. These terms reflect an ambiguity in the quark condensates, inherent to their renormalization in QCD. It is therefore very important to deal with combinations of condensates which are free of this ambiguity. This very same source of ambiguity is also present in lattice simulations at finite $T$. A simple way to get rid of it is to subtract  the $T=0$ contribution. This is the approach followed by the group \cite{Aoki:2009sc} both for condensates and for susceptibilities. A different possibility is to consider the combination $\condtwo-(\hat m)/m_s \conds$ \cite{Cheng:2009zi}, or for individual condensates in the isospin breaking case, $\langle \bar q_i q_i\rangle-(m_i/m_s)\conds$ with $i=u,d$. Another sum rule free of contact ambiguities often used in $T=0$  phenomenology to relate condensate ratios \cite{Gasser:1984gg} is the following  combination:

\begin{eqnarray}
  \Delta_{SR}(T)&\equiv& \frac{\condd_T}{\condu_T}-1+\frac{m_d-m_u}{m_s-\hat m}\left[1-\frac{\conds_T}{\condu_T}\right]=
  \Delta_{SR}(0)+\frac{m_d-m_u}{m_s-\hat m}\frac{1}{F^2}\left[g_1(M_K,T)-g_1(M_\pi,T)\right.\nonumber\\&+&\left.\left(M_K^2-M_\pi^2\right)g_2(M_K,T)\right]-\frac{2Ce^2}{F^4}g_2(M_K,T),
  \label{sumruleT}
\end{eqnarray}
where  $\Od(m_u-m_d)^2$, $\Od(e^4)$ $\Od(e^2(m_u-m_d)^2)$ have been neglected and $\Delta_{SR}(0)$ is given in \cite{Nicola:2010xt} with both sources of isospin breaking contributing at the same order, not only in the chiral counting but also numerically.

\begin{figure}[h]
\includegraphics[scale=.6]{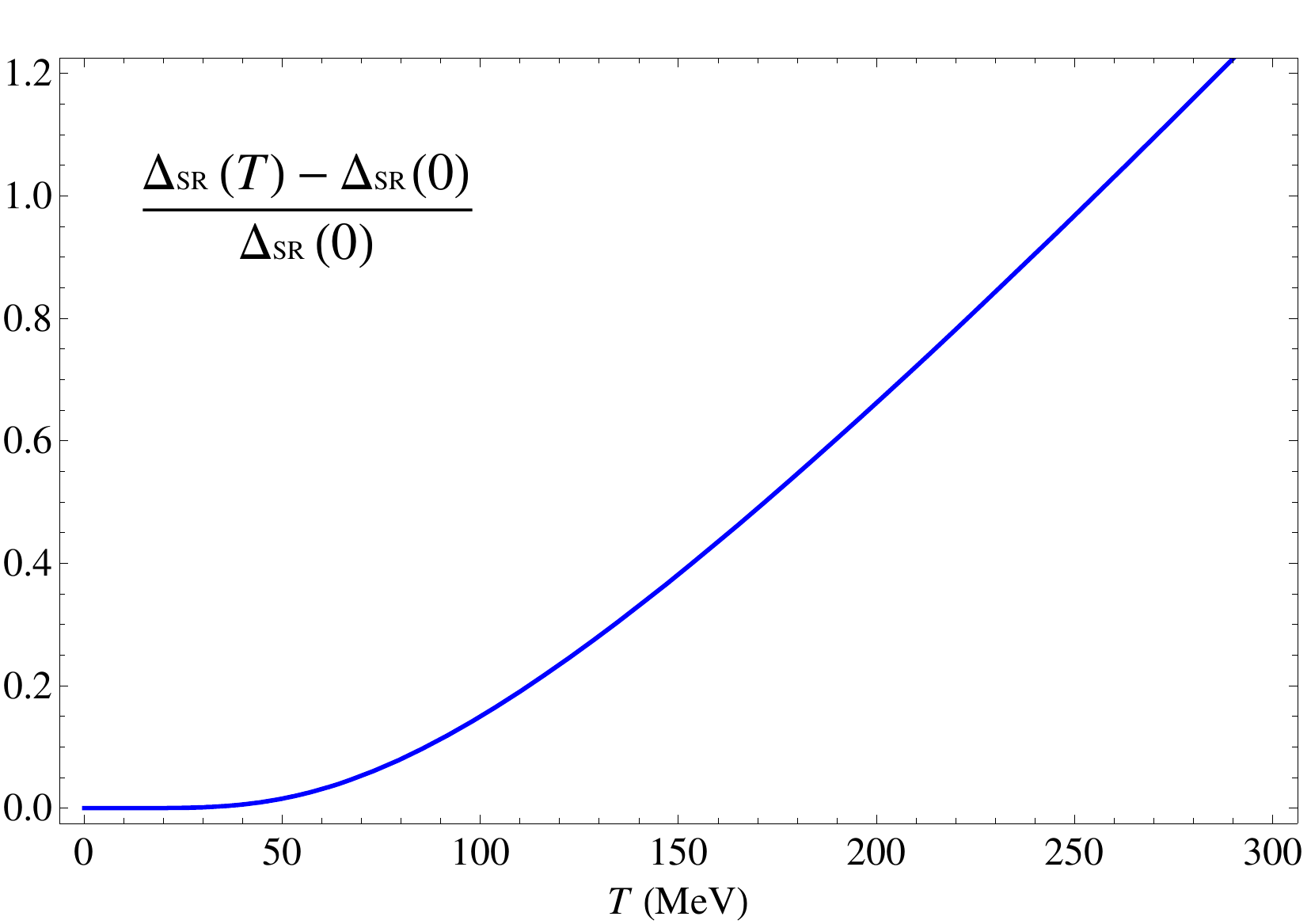}
 \caption{\rm \label{fig:sr} Thermal corrections to the  sum rule relating condensate ratios.}
\end{figure}

We  have seen in section \ref{sec:cond} that the $\condd_T/\condu_T$ ratio receives significant corrections at finite temperature. On the other hand, we expect the strange condensate to vary slowly with $T$, from chiral symmetry breaking due to the strange quark mass. Therefore, we expect that the thermal corrections to this sum rule are also sizable. These  corrections are plotted in Figure \ref{fig:sr} for $K_9^r+K_{10}^r=1/(8\pi^2)$. They become comparable to the $T=0$ sum rule near the critical temperature.

\section{Scalar Susceptibilities and Isospin Breaking}
 \label{sec:susc}

In the isospin-breaking case, the scalar susceptibilities are defined as:

\begin{equation}
\chi_{ij}=-\frac{\partial}{\partial m_i}\langle \bar q_j q_j\rangle_T=\frac{1}{\beta V}\frac{\partial^2}{\partial m_i \partial m_j}\log Z=\int_0^\beta d\tau\int d^3\vec{x}\langle(\bar q_i q_i)(\vec{x},\tau)(\bar q_j q_j)(0,0) \rangle_T-\beta V \langle \bar q_i q_i\rangle_T\langle \bar q_j q_j\rangle_T , \qquad i,j=u,d,s
\label{susdef}
\end{equation}
so that in the light sector, the relevant one concerning chiral restoration, we have three independent scalar susceptibilities  $\chi_{uu}$, $\chi_{dd}$ and $\chi_{ud}=\chi_{du}$.

At this point, it is instructive to recall the definition  of the connected and disconnected parts of the susceptibility. Consider  the isospin limit  with two light identical flavors of mass $\hat m=m_u=m_d$ and $e=0$. There is only one light susceptibility in this case, which can be written as:

\begin{equation}
\chi=-\frac{\partial}{\partial \hat m}\condtwo_T=\frac{1}{\beta V}\frac{\partial^2}{\partial\hat m^2}\log Z=4\chi_{dis}+2\chi_{con},
\label{condistot}
\end{equation}
with:

\begin{eqnarray}
\chi_{dis}&=&\langle \left(\Tr D_l^{-1}\right)^2\rangle_A-\langle \Tr D_l^{-1}\rangle_A^2 ,\label{disdef}\\
\chi_{con}&=&-\langle \Tr D_l^{-2}\rangle_A ,
\label{condef}
\end{eqnarray}
where $D_l=i\!\!\not\!\partial-\hat m$ is the Dirac operator for every light flavor in the QCD lagrangian and $\langle\cdot\rangle_A$ denotes integration over the gluon fields, so that formally $Z=\langle\exp\sum_{j}\Tr\log D_j\rangle_A$ where $j$ runs over flavor and $\Tr$ runs over the space-time, Dirac and color indices.
This separation is important for lattice analysis, as we will discuss below, and reflects the contributions with connected and disconnected quark lines, since $D_l^{-1}$ is the quark propagator. However, when considering the low-energy representation for the partition function and the susceptibilities in terms of GB fields, it is not so simple to separate the connected and disconnected parts if we take the isospin limit from the very beginning. A possible approach to perform such separation  is to work within the partially quenched ChPT framework, as discussed in \cite{DellaMorte:2010aq} for the vacuum polarization. We will however work within the isospin breaking scenario we are considering here, which is very useful for this purpose, as noted first in \cite{Smilga:1995qf} for the susceptibilities and used also in \cite{Juttner:2009yb} for the vacuum polarization. The main point is that for $m_u\neq m_d$:

\begin{equation}
\chi_{ud}=\langle\left(\Tr D_u^{-1}\right)\left(\Tr D_d^{-1}\right)\rangle_A-\langle \Tr D_u^{-1}\rangle_A\langle \Tr D_d^{-1}\rangle_A ,
\end{equation}
so that one has $\chi_{dis}=\lim_{m_u\rightarrow m_d} \chi_{ud}$ and since $\partial_{\hat m}=\partial_{m_u}+\partial_{m_d}$, from (\ref{condistot}) we have also $\chi_{con}=\lim_{m_u\rightarrow m_d}\left[(\chi_{uu}+\chi_{dd})/2-\chi_{ud}\right]$.

Therefore, with this observation in mind, we define in the isospin-breaking regime the following basis of total, connected and disconnected susceptibilities in terms  of the $ij$ basis in (\ref{susdef}):

\begin{eqnarray}
\chi&=&\chi_{uu}+\chi_{dd}+2\chi_{ud} ,\label{chitot}\\
\chi_{con}&=&\frac{1}{2}\left(\chi_{uu}+\chi_{dd}\right)-\chi_{ud},\label{chicon}\\
\chi_{dis}&=&\chi_{ud},
\label{chidis}
\end{eqnarray}
which we can therefore obtain directly from our expressions for the isospin-breaking condensates obtained in the previous section. Observe that none of the $K_i$ dependent terms in the condensates depends on the quark masses and therefore the susceptibilities are independent of the EM LEC.

Note  that according to (\ref{susdef}), $\chi$ in (\ref{chitot}) corresponds to the  correlator of the isosinglet condensate $\condtwo$, the order parameter of chiral restoration, while the connected contribution $\chi_{con}$ is the correlator of the isotriplet $\bar u u-\bar d d$, the order
 parameter for isospin symmetry. A divergence or sudden growth of these susceptibilities would indicate then a phase transition for the corresponding order parameter.

We also remark  that the definitions of the connected and disconnected parts in terms of $uu,dd,ud$ ones are not unique. We could as well have defined  $\chi_{dis}$ as $\alpha(\chi_{uu}-\chi_{dd})+\chi_{ud}$ for arbitrary $\alpha$, which also reduces to the combination (\ref{disdef}) in the isospin limit. We are following the same convention as \cite{Smilga:1995qf}. These formulas can be easily  extended to $N_f$ identical flavors, for which $\chi=N_f\chi_{con}+N_f^2\chi_{dis}$.

In the following we will analyze several aspects related to the above defined susceptibilities in different limits.

\subsection{Sum rule for  EM-like corrections to condensates}
\label{sec:sumrule}

Before studying in detail the different susceptibilities, in this subsection we will relate the EM corrections (and actually any charge-like correction to pion masses) to the condensates, found in section \ref{sec:cond}, with the
 total scalar susceptibility. Consider first the condensate calculated in $SU(2)$ in (\ref{condsu2sum}) and let us define the ratio:

\begin{equation}
r(T)\equiv\frac{\condtwo_T^{e\neq 0}}{\condtwo_T^{e=0}}.
\label{rdef}
\end{equation}

Now note that to one loop, the explicit dependence of the condensate in $e^2$ is only in the $T=0$ part, since the charge dependence in ${\cal L}_2$
 is contained implicitly in the pion mass differences. Therefore, $r(T)-r(0)$ depends on the charge only through the parameter $\delta_\pi\equiv\left(M_{\pi^\pm}^2-M_{\pi^0}^2\right)/M_{\pi^0}^2$, in which we can further expand (for the EM pion mass difference $\delta_\pi\simeq 0.1$). Taking also into account that the condensate is just the sum of the tadpole contributions for the three pions, we can write:

\begin{eqnarray}
r(T)-r(0)&=&-\frac{M_{\pi}^2}{2B_0F^2}\delta_\pi\frac{\partial}{\partial M_{\pi^\pm}^2} \left[\condtwo_T-\condtwo_0\right]+\Od(\delta_\pi^2)+\Od(p^4)\label{rder}\\&=&-\frac{M_{\pi}^2}{6B_0^2F^2}\delta_\pi\frac{\partial}{\partial \hat m} \left[\condtwo_T-\condtwo_0\right]+\Od(\delta_\pi^2)+\Od(p^4),
\end{eqnarray}
which, from the susceptibility definition in (\ref{chitot}) can be written, to this order, as:

\begin{equation}
r(T)-r(0)=\frac{2}{3}\frac{\hat m^2}{M_\pi^2F^2}\delta_\pi\left[\chi(T)-\chi(0)\right].
\label{rsumrulegen}
\end{equation}

This sum rule relates then  pion mass deviations in the condensate with the total scalar susceptibility. Note that the above result is written only in terms of the quark mass, the pion mass and decay constant and the charged-neutral mass difference, without specifying if the latter is of electromagnetic origin.  It states that, even though the mass deviation $\delta_\pi$  may be small, the  corrections to the condensate may be amplified near the phase transition, where the susceptibility is maximum, if such transition is sufficiently strong. Actually, the quantity proportional to $\delta_\pi$ on the right hand side of (\ref{rsumrulegen}) is directly measurable on the lattice \cite{Aoki:2009sc}.

For the case for the electromagnetic mass difference in $SU(2)$ discussed in section \ref{sec:form}, we have $\delta_\pi M_{\pi^0}^2=2Ce^2/F^2$ and:

\begin{equation}
r(T)-r(0)= \frac{2Ce^2}{F^4}g_2(M_{\pi^0},T) +\Od(e^4),
\label{rsu2}
\end{equation}
with:

\begin{equation}
r(0)=1+e^2\mathcal{K}_2^r(\mu)-\frac{4Ce^2}{F^4}\nu_{\pi^0},
\label{r0su2}
\end{equation}
and
\begin{eqnarray}
\nu_i&=&F^2\frac{d}{dM_i^2}\mu_i=\frac{1}{32\pi^2}\left[1+\log\frac{M_i^2}{\mu^2}\right],\nonumber\\
g_2(M,T)&=&-\frac{dg_1(M,T)}{dM^2}=\frac{1}{4\pi^2}\int_0^\infty dp \frac{1}{E_p} \frac{1}{e^{\beta E_p}-1}.
\label{nug2def}
\end{eqnarray}

Note that  $r(T)$ is finite, scale-independent and also independent of  the $e=0$ LEC $l_3,h_1,h_3$. In particular,  it is free of the contact-terms ambiguity, which makes it a  quantity  suitable for physical predictions. It is also independent of $B_0$, unlike the individual quark condensates, which have only physical meaning when multiplied by the appropriate quark masses, since the $m_i B_0$ products give meson masses. In addition,  the dependence with the EM LEC disappears in the difference $r(T)-r(0)$, which is the quantity directly related to the susceptibility
through (\ref{rsumrulegen}).

The above relation can also be explored for $SU(3)$. However, the
connection  with the susceptibility is not direct in that
case. The $e^2$ dependence of $\condtwo_T-\condtwo_0$ enters now
through $M_{\pi^\pm}$ and $M_{K^\pm}$. However, the condensates
(\ref{condsu3sum})-(\ref{condsu3str}) depend on the light quark
mass through all the meson masses $M_\pi,M_K,M_\eta$. The
result is that $r(T)-r(0)$ can be expressed as the susceptibility
term in (\ref{rsumrulegen}) plus a linear combination of $\partial
(\conds_T-\conds_0)/\partial m$ and $\partial
(\conds_T-\conds_0)/\partial m_s$, to this chiral order and
neglecting $\Od(e^4)$ and $\Od(m_u-m_d)$ isospin-breaking corrections in the right hand side.
Since the strange quark condensate  has a much weaker dependence
on temperature than the light one (or equivalently, we can approximately neglect the thermal functions evaluated on kaon and eta masses) we expect the $T$ behaviour of
$r(T)-r(0)$ to be dominated by the light scalar susceptibility
also in the $SU(3)$ case and therefore the sum rule (\ref{rsumrulegen}) should hold approximately.  In this case we have to one loop:

\begin{eqnarray*}
r(T)^{SU(3)}-r(0)^{SU(3)}&=&1+\frac{C e^2}{F^4}\left[2g_2(M_{\pi^\pm},T)+g_2(M_{K^\pm},T)\right]+\Od(e^4),\nonumber\\
r(0)^{SU(3)}&=&1+e^2 \mathcal{K}_{3+}^r (\mu)-\frac{2C e^2}{F^4}\left[2\nu_{\pi^\pm}+\nu_{K^\pm}\right]+\Od(e^4),
\label{rsu3}
\end{eqnarray*}
where the expansion in $e^2$ to leading order allows to express the result in terms of the $\pi^\pm$ and $K^\pm$ masses.
As in $SU(2)$, $r(T)$ is finite, scale-independent and independent of the $e=0$ LEC, so that it is free of contact ambiguities.

We compare the above expression with the susceptibility in the $SU(3)$ case. As it will become clear in section \ref{sec:sustempmass}, the corrections to the total susceptibility $\chi$ from both sources of isospin breaking are small. Actually, we will see that the NLO correction in the QCD breaking is $\Od(m_u-m_d)^2$. The one-loop result is:

\begin{eqnarray}
  \frac{\chi(T)-\chi(0)}{B_0^2}&=&\frac{4\hat m^2}{M_\pi^4}\left[\chi(T)-\chi(0)\right]=2\left[3g_2(M_\pi,T)+g_2(M_K,T)+\frac{1}{9}g_2(M_\eta,T)\right]
  +\Od\left(p^2\right)+\Od(e^2)+\Od\left(\frac{m_d-m_u}{m_s}\right)^2,\nonumber\\\label{sustotsu3T}\\
  \frac{\chi (0)}{B_0^2}&=&\frac{4\hat m^2}{M_\pi^4}\chi (0)=16\left[8L_6^r(\mu)+2L_8^r(\mu)+H_2^r(\mu)\right]-4\left[3\nu_\pi+\nu_K+\frac{1}{9}\nu_\eta\right]+\Od\left(p^2\right)+\Od(e^2)+\Od\left(\frac{m_d-m_u}{m_s}\right)^2
  \nonumber .\\\label{sustotsu3zero}
\end{eqnarray}

\begin{figure}[h]
\includegraphics[scale=.6]{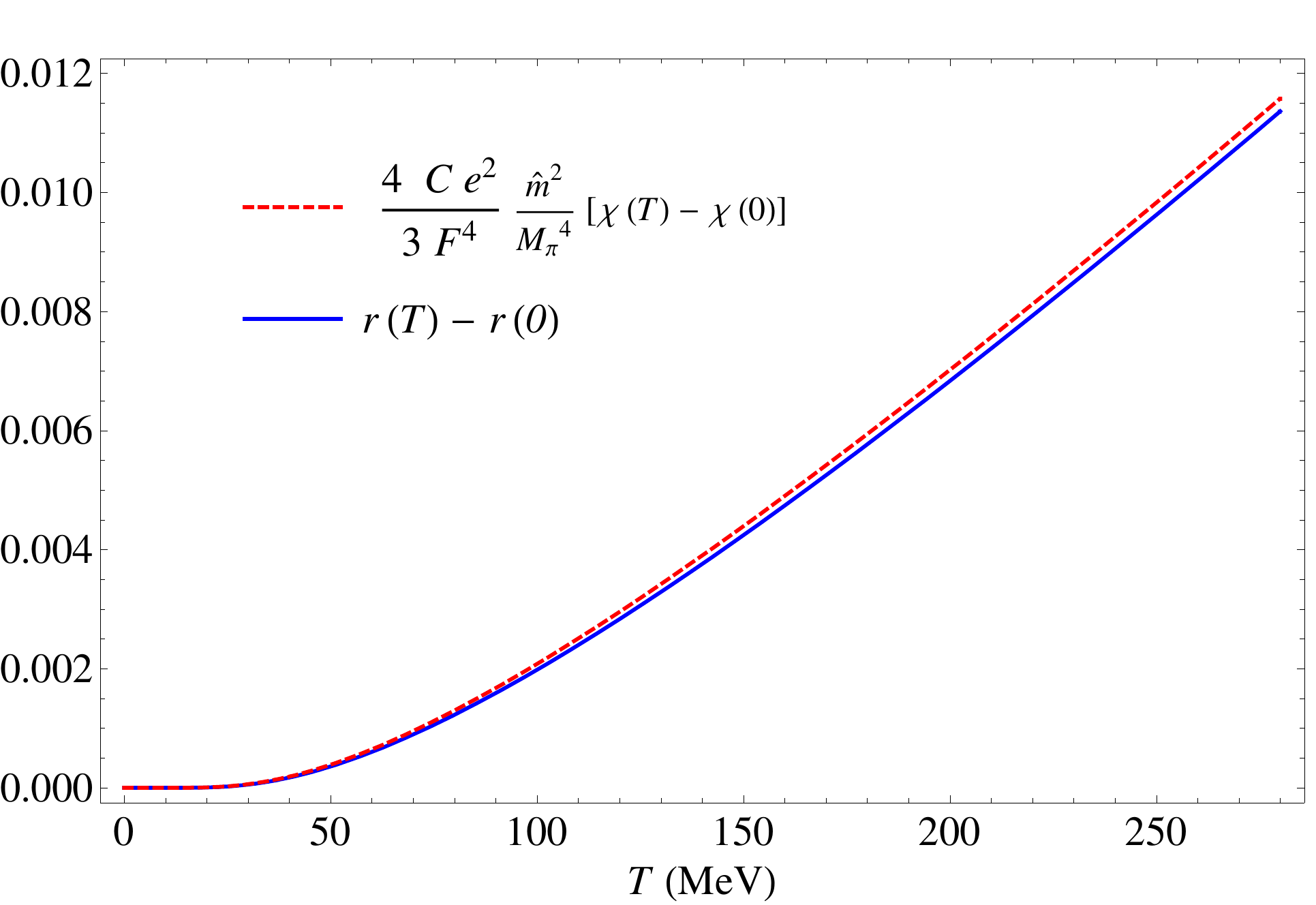}
 \caption{\rm \label{fig:r} The $r(T)$ function in $SU(3)$  encoding the EM corrections of the quark condensate. We compare it with the normalized light susceptibility for physical quark and meson masses. For these values, $r(0)\simeq 1.01$.}
\end{figure}

The two quantities are compared in Figure \ref{fig:r}. The deviations between them are negligible for the range of relevant temperatures.
 Therefore, although for physical masses the electromagnetic corrections are relatively small, they grow with the susceptibility, which is a model-independent prediction. For comparison, taking the value of $(\hat m^2)/(M_\pi^4)\left[\chi(T_c)-\chi(0)\right]$ from the  lattice simulations in \cite{Aoki:2009sc} for 2+1 flavors with the lattice $T_c$ value gives $r(T_c)-r(0)\simeq0.013$, not far from the higher temperature values in Figure \ref{fig:r}, although the ChPT curve cannot reproduce the susceptibility peak, only the low and moderate $T$ behaviour. These small EM corrections for the condensate are in accordance with our simple estimates made in section \ref{sec:cond} and translate into a few MeV difference in the determination of the critical temperature from the order parameter.

The sum rule (\ref{rsumrulegen}) has another interesting consequence, regarding lattice simulations. In the staggered fermion lattice formalism, the need to introduce four different copies (tastes) for every quark flavor leads to the so called taste violation \cite{Lee:1999zxa,Aoki:2009sc,DeTar:2009ef}. This is a lattice artifact which in some aspects is similar to the isospin or flavor violations we are analyzing here. The new tastes enlarge the chiral symmetry group  to $SU(4N_f)\times SU(4N_f)$, producing then 15 pseudo-Goldstone bosons plus one massive state ($\eta'$-like) for every quark flavor. All these new meson states become degenerate in the continuum limit, where taking the fourth root of the Dirac fermion determinant is enough to remove all the spurious copies. However, for finite lattice spacing $a$, the tree-level masses of those  states receive $\Od(a^2)$ contributions, which break explicitly the chiral group in the lagrangian,
only one Goldstone boson remaining massless in the chiral limit leaving then a residual
$O(2)$ or $U(1)$ symmetry. The mechanism is similar to the electric charge one we are analyzing here, by which the charged states receive $\Od(e^2)$ corrections and the $U(1)$ EM symmetry remains. In fact, for the staggered case one can construct a generalized chiral lagrangian  including all possible terms compatible with the new symmetry. This is called staggered Chiral Perturbation Theory \cite{Lee:1999zxa,Aubin:2003mg}. Among the new terms one recognizes contributions of the form $\Tr[\xi U \xi U^\dagger]$ with $\xi$ a given combination of $SU(4)$ generators, i.e., like the charge  term in (\ref{L2}) in $SU(3)$. Obviously, the staggered case includes other operator structure and the spectrum of states is  more complicated. However, we can use the sum rule (\ref{rsumrulegen})  to estimate roughly the expected differences between  the lattice staggered condensate and the continuum  one, considering the lightest states. For fine enough lattices, one has   $\delta_\pi^a=(M_{\pi,a}^2-M_\pi^2)/M_\pi^2\simeq c a^2$ for the lightest  tastes of squared mass  $M_{\pi,a}^2$ \cite{Aoki:2009sc}, where  from the two smallest lattices in \cite{Aoki:2009sc} we get $c\simeq 140$ fm$^{-2}$.  With this $\delta_\pi^a$ we can then use  (\ref{rsumrulegen}) to estimate $r^a(T)-r^0(0)$. Consequently, we expect the larger errors coming from this taste violation effect to appear near $T_c$. That is indeed the case when we compare lattices of decreasing temporal extent $N_t=a/T$ for the condensate data given in \cite{Aoki:2009sc}.
More quantitatively, taking also the susceptibility values of \cite{Aoki:2009sc}, we get $r^a(T_c)-r^a(0)\simeq 0.07 c a^2$. Estimating the $T=0$ part using (\ref{r0su2}) with $e^2\rightarrow \delta_\pi^a M_\pi^2 F^2/(2C)$, we get  a  relative correction for the condensates near $T_c$ with respect to the continuum of about 20\% for the $N_t=12$ data in   \cite{Aoki:2009sc} and about 12\% for the $N_t=16$ ones in \cite{Borsanyi:2010bp}. Following the same idea, we get a  relative difference between the $N_t=10$ and $N_t=12$ lattices of around 8\% near $T_c$, which is actually in good agreement with  lattice data  \cite{Aoki:2009sc}. A direct translation into an error for the critical temperature is not easy to obtain. If we simply extrapolate the one-loop chiral limit  expression   $\condtwo_T=\condtwo_0(1-T^2/T_c^2)$, writing the l.h.s of (\ref{rsumrulegen}) in the one-loop equivalent form (\ref{rder}), we get a very rough estimate $\Delta T_c \simeq$ 10 MeV for $N_t=12$ with respect to the continuum, although the chiral limit is not always numerically accurate, as we will actually see in the next section.

Estimating taste violation effects is important,  since they are one of the main sources of the discrepancies between different lattice groups for the determination of the critical temperature. An important effort has been made over recent years to minimize these effects, not only by considering finer  lattices, but also by introducing lattice actions where taste symmetry is reduced \cite{Borsanyi:2010bp}.

\subsection{Temperature and mass dependence of connected and disconnected susceptibilities. Relation with chiral restoration and lattice analysis}
\label{sec:sustempmass}

The behaviour of condensates and susceptibilities with  temperature and quark masses is crucial in order to understand
 the nature of the chiral phase transition when approaching the chiral region $(m_q,T)\rightarrow(0^+,T_c)$. Lattice simulations have addressed the question of how those quantities scale with $m_q$ and $T$  until
  very recently \cite{Ejiri:2009ac}.  An essential part of this program concerns the scaling of the connected and disconnected parts of the scalar susceptibility. The disconnected piece is given in terms of closed quark lines and is therefore directly related to $\condtwo$ and expected to
  be  sensitive to chiral restoration.  Actually, near the chiral limit, i.e., the infrared (IR) behaviour, it is known to scale as $\chi_{dis}^{IR}\sim \log M_\pi^2$ for $T=0$ and $\chi_{dis}^{IR}\sim T/M_\pi$ at finite temperature \cite{Smilga:1995qf}. The infrared contribution is then controlled by the
  GB loop contributions. The situation is not so clear for the connected part, since its infrared divergent piece is proportional to $n_f^2-4$, with $n_f$ the number of identical light flavors \cite{Smilga:1995qf} and therefore vanishes for $n_f=2$. However, its IR finite part contributes in the physical case of massive pions and is actually an important difference between QCD and $O(N)$ models in which the lattice scaling fits are based \cite{Ejiri:2009ac}.
   Besides, the connected contribution receives important ``false" GB-like corrections coming from taste violation \cite{Ejiri:2009ac,Unger:2009zz}. It is therefore important for lattice studies to provide the continuum result for the disconnected and connected susceptibilities in the physical case of 2+1 flavors and massive pions.

 Our present ChPT one-loop analysis allows to obtain a model-independent prediction  for the low temperature and small mass behaviour of the susceptibilities. The inclusion of isospin breaking effects is crucial. In fact, it will be useful for the following discussion to note that:

 \begin{eqnarray}
 \chi_{con}&=&\frac{1}{2}\left(\chi_{uu}-\chi_{ud}\right)+\frac{1}{2}\left(\chi_{dd}-\chi_{ud}\right)=-\frac{1}{2}\partial_{m_\delta}\conddif,\nonumber\\
 \chi_{dis}&=&-\frac{1}{4}\left[\partial_{\hat m}\condsum-2\partial_{m_\delta}\conddif\right],
 \label{susconmdelta}
 \end{eqnarray}
 with $m_\delta=(m_u-m_d)/2$, so that $\chi_{con}$ comes only from the condensate difference in (\ref{condsu3dif}) and  its leading order is obtained from the linear terms in $m_u-m_d$.

 Now, from our considerations in section \ref{sec:cond}, if we neglect for the moment the charge corrections and we take the quark mass derivatives in (\ref{condsu3sum}) we have, to leading order in $m_\delta$, $\chi_{uu}\simeq \chi_{dd}$, $\chi_{uu}+\chi_{ud}\simeq\chi/2$ and $\chi_{uu}-\chi_{ud}\simeq\chi_{con}$ where the leading terms $\chi$, $\chi_{con}$ and $\chi_{dis}=\chi/4-\chi_{con}/2$ are $\Od(1)$ in the $m_\delta$ counting and are given in $SU(3)$ by (\ref{sustotsu3T})-(\ref{sustotsu3zero}) and:

\begin{eqnarray}
\frac{\chi_{dis}(T)-\chi_{dis}(0)}{B_0^2}&=&\frac{1}{18}\left[27g_2(M_\pi,T)+g_2(M_\eta,T)\right]-\frac{g_1(M_\pi,T)-g_1(M_\eta,T)}{3(M_\eta^2-M_\pi^2)}
+\Od(p^2)+\Od\left(\frac{m_d-m_u}{m_s}\right)^2+\Od(e^2), \nonumber\\
\label{chidisTlo}\\
\frac{\chi_{dis}(0)}{B_0^2}&=&\frac{1}{9}\left(288 L_6^r(\mu)-\nu_\eta-27\nu_\pi\right)-\frac{2F^2\left(\mu_\pi-\mu_\eta\right)}{3(M_\eta^2-M_\pi^2)}+\Od(p^2)+\Od\left(\frac{m_d-m_u}{m_s}\right)^2+\Od(e^2),
\label{chidiszerolo}\\
\frac{\chi_{con}(T)-\chi_{con}(0)}{B_0^2}&=&g_2(M_K,T)+\frac{2\left[g_1(M_\pi,T)-g_1(M_\eta,T)\right]}{3(M_\eta^2-M_\pi^2)}
+\Od(p^2)+\Od\left(\frac{m_d-m_u}{m_s}\right)^2+\Od(e^2),
\label{chiconTlo}\\
\frac{\chi_{con}(0)}{B_0^2}&=&2\left(4H_2^r(\mu)+8L_8^r(\mu)-\nu_K\right)
+\frac{4F^2\left(\mu_\pi-\mu_\eta\right)}{3(M_\eta^2-M_\pi^2)}+\Od(p^2)+\Od\left(\frac{m_d-m_u}{m_s}\right)^2+\Od(e^2).
\label{chiconzerolo}
\end{eqnarray}

Several remarks are in order from the previous expressions. First, we emphasize that all of them are finite and scale-independent, which can be explicitly checked from the scale dependence of the LEC \cite{Nicola:2010xt}. We also note that, unlike the disconnected part, the connected susceptibility does not receive contributions from the  mass derivative of pion tadpoles, i.e $\nu_\pi$ or $g_2(M_\pi,T)$. These turn out to be the dominant ones in the chiral limit (see below) and this is what we expected from our previous discussion on the infrared behaviour.  Thus, we identify the  difference $\mu_\pi-\mu_\eta$  in  (\ref{chiconzerolo}) and the corresponding $g_1$ one in (\ref{chiconTlo}) as the contribution of $\pi^0\eta$ mixing, while the $\nu_K,g_2(M_K)$ terms come from the expansion of the kaon contribution in the r.h.s. of (\ref{condsu3dif}) around the isospin limit. From (\ref{susconmdelta}), the disconnected part receives in addition a contribution from the sum (\ref{condsu3sum}) and hence it incorporates the $\nu_\pi$, $g_2(M_\pi)$ terms. Note also that the $T=0$ part of $\chi_{con}$ depends on the contact term $H_2$, while that dependence cancels in $\chi_{dis}$. This means that only quantities such as $\chi_{con}(T)-\chi_{con}(0)$ can be unambiguously determined, similarly as the quark condensate case (see our discussion at the end of section \ref{sec:cond}). This is a relevant comment for lattice evaluations of this quantity. In fact, as a consequence of the vanishing pion terms, we see that $\chi_{con}(T)-\chi_{con}(0)$ vanishes formally in the $m_s\rightarrow\infty$ limit, recovering the pure $SU(2)$ result  that we would get from (\ref{condsu2dif}), which holds also for the other  susceptibilities taking into account the conversion between the $SU(2)$ and $SU(3)$ LEC \cite{Nicola:2010xt} (see below).

Regarding the expansion in $(m_u-m_d)$, it can be seen from (\ref{susconmdelta}), the condensate expressions (\ref{condsu3sum}), (\ref{condsu3dif}) and the meson masses and mixing angle dependence on $\hat m$ and $m_\delta$, that the linear order in $m_\delta$ cancels both in the connected and the disconnected parts and so we have  written in the previous expressions and  in the total susceptibility (\ref{sustotsu3T})-(\ref{sustotsu3zero}). It is important to remark that the $\Od(m_\delta/m_s)^2\sim\Od(\varepsilon)^2$ corrections contain ''tadpole mass derivative" terms $\nu_\pi,g_2(M_\pi)$ in both $\chi_{con}$ and $\chi_{dis}$. However, an important  difference between them  is that those IR dominant terms do not appear to leading order in the connected contribution. In fact, since $\nu_{\pi^0}=\nu_{\pi}+(\partial\nu_\pi/\partial M_\pi^2)\Od(m_\delta/m_s)^2$, where the subscript $\pi$ indicates just the $m_u=m_d$ and $e^2=0$ pion, the disconnected part receives an $\Od(m_\delta/m_s)^2$ proportional to the second derivative of the pion tadpole and hence more IR divergent, and so on for the thermal part $g_2(M_{\pi^0},T)$. This contribution is not present in the connected part to that order. Thus,  we expect the isospin-breaking corrections to be larger in the disconnected than in the connected susceptibility.

Another pertinent comment is that we have been able to obtain the leading order for $\chi_{con}$ and $\chi_{dis}$ only after considering properly all the $m_u\neq m_d$ contributions and then taking the $m_u=m_d$ limit. However, one can be led to misleading results by setting $m_u=m_d$ from the very  beginning. For instance, for two equal masses one could think naively that $\chi_{uu}=\chi_{ud}=\chi_{dd}= \chi/4$, from the definition (\ref{chitot}). However, from our previous analysis we see that in the isospin limit what we get actually is $\chi_{ud}=\chi/4-\chi_{con}/2$  and $\chi_{uu}=\chi_{dd}=\chi/4+\chi_{con}/2$ with $\chi_{con}$ given in (\ref{chiconTlo})-(\ref{chiconzerolo}) i.e, not vanishing for $m_u=m_d$ and  physical $m_s$, although formally suppressed in the $m_s\rightarrow\infty$ limit. Thus, in the isospin limit taking just one flavor susceptibility and multiplying by four does not give the total scalar susceptibility, which should be  obtained instead by considering the derivative of the full sum of $u$ and $d$ condensates as given by (\ref{chitot}). Note that this correction does not affect to condensates, for which the $\condu$ and $\condd$ difference vanishes for $m_u=m_d$ and $e^2=0$ and therefore $\condtwo=2\condu=2\condd$ in the isospin limit. This comment may be relevant for certain lattice analysis, where working within the ``one-flavor" equivalent framework is often done, since in that way it is easy to discuss for instance the taste-breaking effect.  This observation could then help to explain the worse $O(N)$ scaling properties of the susceptibility with respect to those of the condensate  \cite{Ejiri:2009ac}. In that work, the lattice data for the susceptibility scaling function in the 2+1 case suffer from a sizable increase as the strange quark mass is decreased relative to the light one. That increase is not seen in the quark condensate data and could be due partly to the definition used as we have just explained, since the positive term proportional to $\chi_{con}$ increases with $1/m_s$, from (\ref{chiconTlo})-(\ref{chiconzerolo}). This is not the only effect that may cause this ``wrong" scaling, since, as pointed out in \cite{Ejiri:2009ac,Unger:2009zz} taste-breaking induces an artificial infrared pion contribution in $\chi_{con}$ which is not present in the continuum, as our above expressions show. What we are pointing out here is that  the ``one-flavor" $\chi_{uu}$ and $\chi_{dd}$ are sensitive to isospin-breaking terms even for $m_u=m_d$ and in the continuum, unlike considering for instance the total susceptibility $\chi\sim \chi_{uu}+\chi_{ud}$ in (\ref{chitot}) for which the $\chi_{con}$ term cancels. In addition, as we will see below, $\chi_{con}$ is dominated numerically by its $T=0$ part, which would explain why the anomalous scaling is reduced for the subtracted susceptibility.

The charge corrections to susceptibilities in our previous expressions (\ref{sustotsu3T})-(\ref{chiconzerolo}) arise only through the mass of the charged mesons $\pi^\pm$ and $K^\pm$. Thus, including the charge amounts to replace  $3\nu_\pi\rightarrow \nu_{\pi^0}+2\nu_{\pi^\pm}$  in (\ref{sustotsu3zero})  and (\ref{chidiszerolo}) and $2\nu_K\rightarrow \nu_K+\nu_{K^\pm}$ in  (\ref{sustotsu3zero})  and (\ref{chiconzerolo}) and so on for the thermal parts $3g_2(M_\pi,T)\rightarrow g_2(M_{\pi^0},T)+2g_2(M_{\pi^\pm},T)$ and $2g_2(M_K,T)\rightarrow g_2(M_K,T)+g_2(M_{K^\pm},T)$. Although for physical values of the electric charge and masses, these represent small perturbative corrections, the fact that near the chiral limit the coefficient of the IR-dominant $\nu_\pi$ and $g_2(M_\pi,T)$ reduces in $1/3$ for $e^2\neq 0$ is the reflection in the scalar susceptibility of the behaviour of the condensate in terms of $\delta_\pi$ corrections to the masses analyzed in section \ref{sec:sumrule}. Thus, when the mass corrections $\delta_\pi$ become sizable, as in the staggered lattice formalism, the susceptibility is reduced by that factor, which eventually would imply that the transition peak or maximum is displaced to a higher temperature, consistent with our analysis in the previous section about the increasing of $T_c$ in the condensates. Recall that one cannot just expand the pion terms in $e^2$ and then take the chiral limit, since that expansion assumes that the charge part of $M_{\pi^\pm}$ is small compared to the quark mass one.

Before continuing, we also remark that our above results  are compatible with the recent observation \cite{Unger:2009zz} that the connected and disconnected susceptibilities can be inferred from  the zero momentum limit of the $a_0$ and $f_0$ correlators calculated previously  in staggered ChPT \cite{Bernard:2007qf}. The motivation of those works is precisely to estimate the contribution of heavy pion-like tastes to the IR part of $\chi_{con}$, which could mask the scaling behaviour. The continuum limit of the results in \cite{Unger:2009zz} reveals the same $\pi,K,\eta$ loop contributions as in our expressions (\ref{chidisTlo})-(\ref{chiconzerolo}). We provide the full ChPT result, including the LEC contribution necessary to guarantee the finiteness and scale-independency of the results, as well as the analysis of the higher order corrections in isospin breaking.

As discussed above, the behaviour of the susceptibilities near the chiral limit (IR regime) is very illuminating regarding their approach to chiral restoration within the $O(4)$ or $SU(2)$ pattern in the continuum, i.e., without taste breaking effects. Let us  consider this regime first for $T=0$ and only for the leading order terms in the isospin expansion in (\ref{chidiszerolo})-(\ref{chiconTlo}), i.e., we set $m_u=m_d=\hat m$ and $e^2=0$ in the $T=0$ susceptibilities and consider $\hat m \ll m_s$. We denote by a superscript ``IR"  the nonvanishing terms in that limit:

\begin{eqnarray}
\frac{\chi_{dis}^{IR}(T=0)}{B_0^2}&=&-\frac{3}{32\pi^2} \log\frac{M_\pi^2}{\mu^2}+32L_6^r(\mu)+\frac{1}{288\pi^2}\left(-28+5\log\frac{M_\eta^2}{\mu^2}\right),\label{chidiszeroir}\\
\frac{\chi_{con}^{IR}(T=0)}{B_0^2}&=&8\left[H_2^r(\mu)+2L_8^r(\mu)\right]-\frac{1}{16\pi^2}\left(1+\log\frac{M_K^2}{\mu^2}
+\frac{2}{3}\log\frac{M_\eta^2}{\mu^2}\right).\label{chiconzeroir}
\end{eqnarray}

The IR divergent $\log M_\pi^2$ term in (\ref{chidiszeroir}) coincides with the one obtained in \cite{Smilga:1995qf}, where a cutoff regularization was used. Multiplied by 4, this is also the IR dominant part in the total susceptibility $\chi$. In addition to that term, we obtain here the regular part, not IR divergent but not vanishing in the chiral limit, which provides the dependence with the LEC and together with the logarithm gives the consistent scale-independent ChPT prediction. As for the connected part, again only the IR divergent contribution is given in \cite{Smilga:1995qf}, which as commented  above turns out to vanish exactly for two light fermions. Here we also give the regular contribution, also scale-independent, which unlike the disconnected part depends on the contact term $H_2^r$. As an interesting consistency check, we can recover the $SU(2)$ limit from the above expressions, using  the conversion between the LEC of $SU(2)$ and $SU(3)$ \cite{Gasser:1984gg} $l_3^r+h_1^r-h_3=16L_6^r+5\nu_\eta/18-1/(96\pi^2)$ and  $h_3^r=4L_8^r+2H_2^r-\nu_K/2-\nu_\eta/3+1/(96\pi^2)$, so that $\chi_{dis}^{SU(2)}=-3\nu_\pi+2(l_3^r+h_1^r-h_3)$ and $\chi_{con}^{SU(2)}=4h_3^r$ , which is the same result that we would have starting directly from the $SU(2)$ expressions in (\ref{condsu2sum}) and (\ref{condsu2dif}).

Let us consider now the dominant IR thermal contribution in the 2+1 flavor case, i.e, apart from $\hat m \ll m_s$ we also consider temperatures $M_\pi \ll T \ll M_K$, so that we neglect all the Boltzmann exponentials $\exp(-M_{K,\eta}/T)$  and expand $g_1(M_\pi,T)=\frac{T^2}{12}[1-3M_\pi/(\pi T)+\Od(M_\pi^2\log M_\pi^2)]$ and $g_2(M_\pi,T)=T/(8\pi M_\pi)+\Od(\log M_\pi^2)$ \cite{Gerber:1988tt}. Thus, we get:

\begin{eqnarray}
\frac{\left[\chi_{dis}(T)-\chi_{dis}(0)\right]^{IR}}
{B_0^2}&=&\frac{3T}{16\pi M_\pi},\label{chidisTir}\\
\frac{\left[\chi_{con}(T)-\chi_{con}(0)\right]^{IR}}{B_0^2}&=&\frac{T^2}{18M_\eta^2}.\label{chiconTir}
\end{eqnarray}

The disconnected part (\ref{chidisTir}) is again the one
 obtained in \cite{Smilga:1995qf} in the IR limit. It diverges more strongly than
 the $T=0$ contribution in (\ref{chidiszeroir}) in this limit, revealing its critical behaviour. The growth with $T$ is linear over the GB mass scale. Recall that, apart from the heavy masses thermal exponentials, we are neglecting also $\log M_\pi$ terms in the disconnected part (\ref{chidisTir}). The situation is completely different for the connected contribution, which is regular, albeit not vanishing, in the chiral limit. The quadratically growing term in (\ref{chiconTir}) survives for $M_\pi\rightarrow 0$  against neglected $\Od(M_\pi)$ and is dominant over $\exp(-M_{K,\eta}/T)$. It vanishes formally as $m_s\rightarrow\infty$, recovering the $SU(2)$ limit. For physical masses though, an specific and model-independent difference between the $N_f=2$ and $N_f=2+1$  cases is the (soft) temperature dependence of the connected susceptibility, the scale that controls its growth  being $M_\eta^2$ instead of the $M_\pi^2$ of the connected part. This is  a consequence of  $\chi_{dis}$ measuring the fluctuations of the chiral restoration order parameter, while $\chi_{con}$ is related to those of the isospin-breaking one, i.e., $\conddif$, which as we have seen in section \ref{sec:cond} increases moderately. Note that near the chiral limit we could as well have written the $T^2$ term divided by $M_K^2$ just by changing the multiplying factor. Keeping  $M_\eta^2$ reminds its $\pi^0\eta$ origin, as it is clearly seen in the original expressions in (\ref{chiconTlo})-(\ref{chiconzerolo}).

\begin{figure}[h]
\includegraphics[scale=.52]{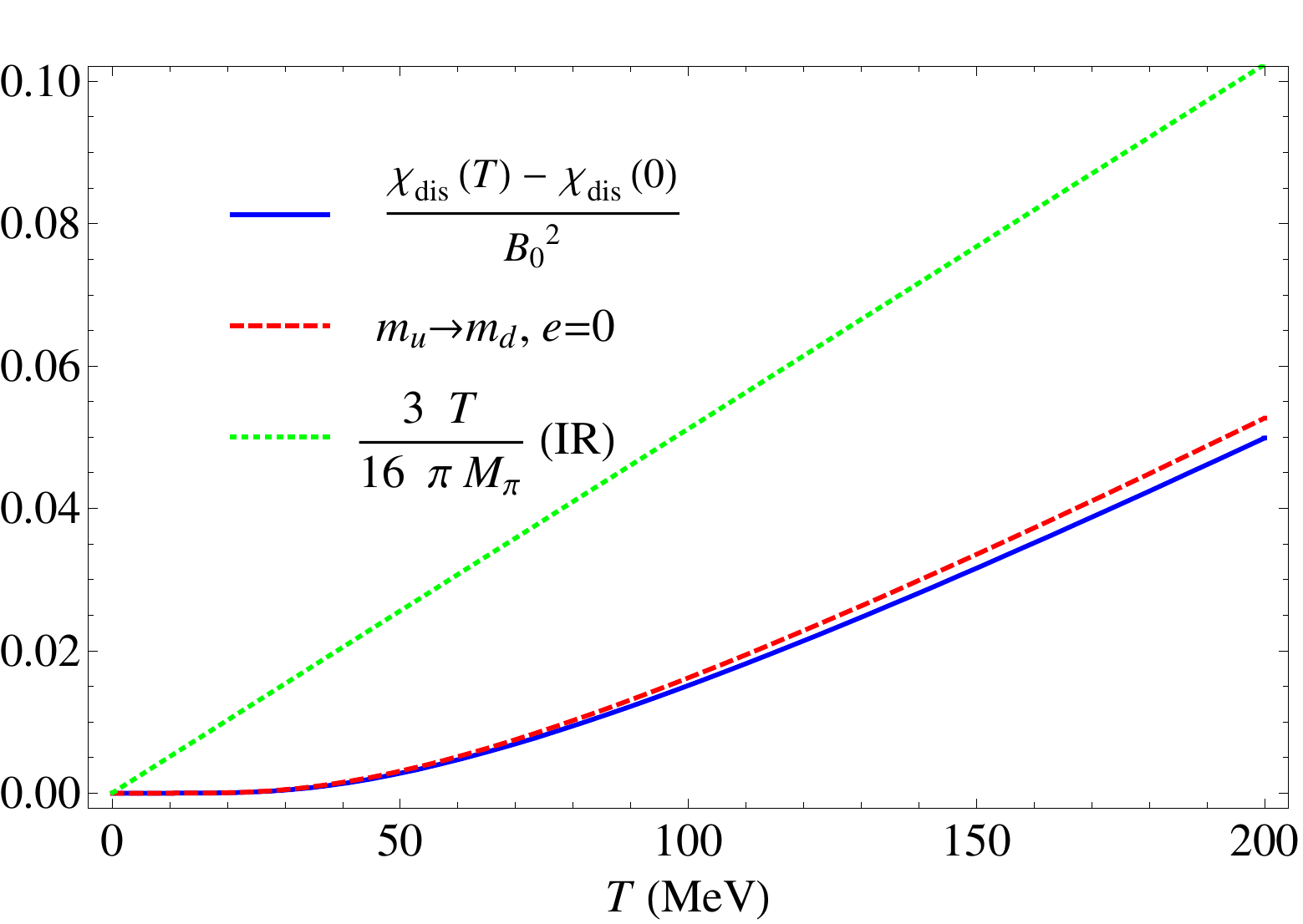}\hspace*{0.2cm}\includegraphics[scale=.5]{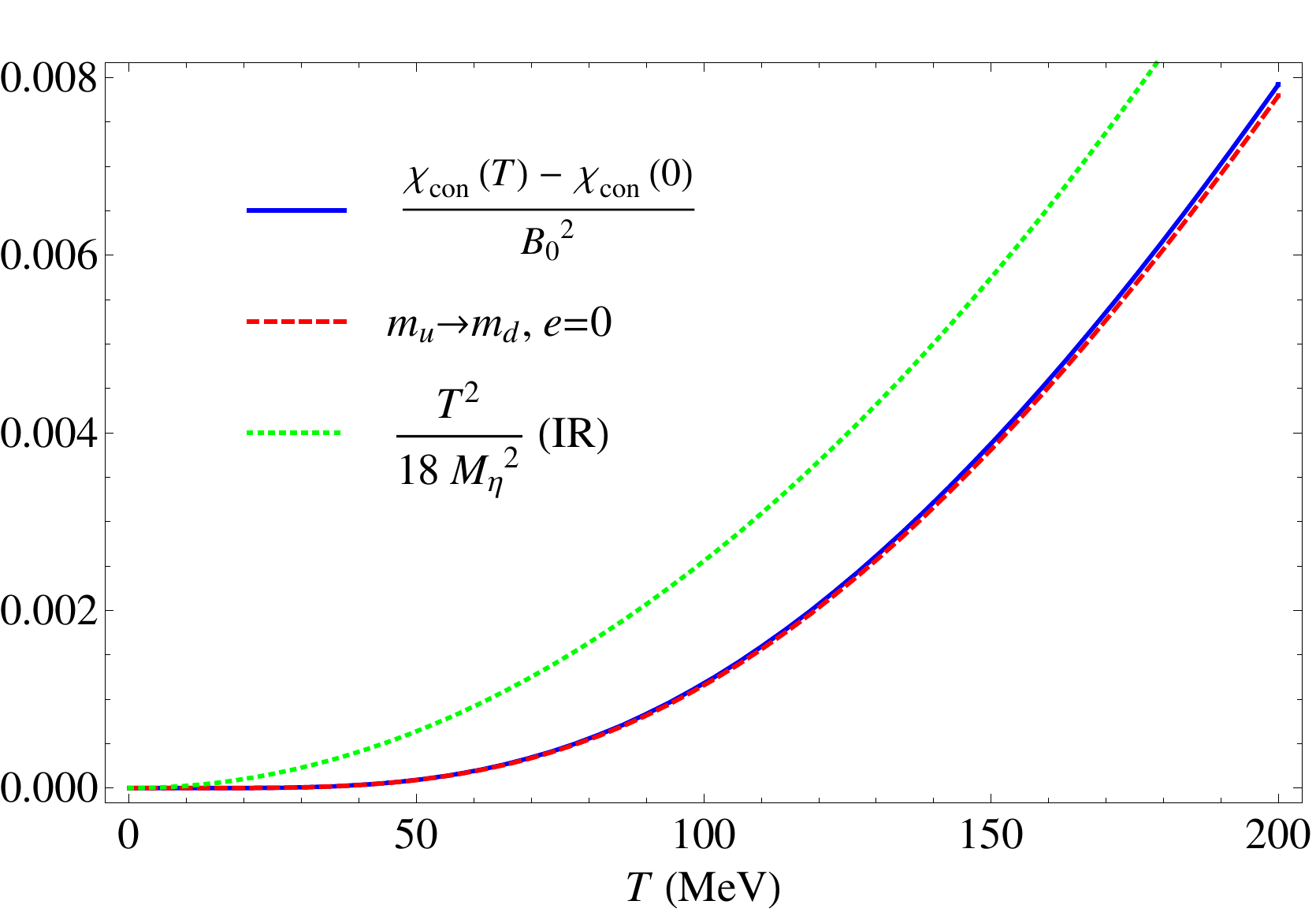}
\includegraphics[scale=.52]{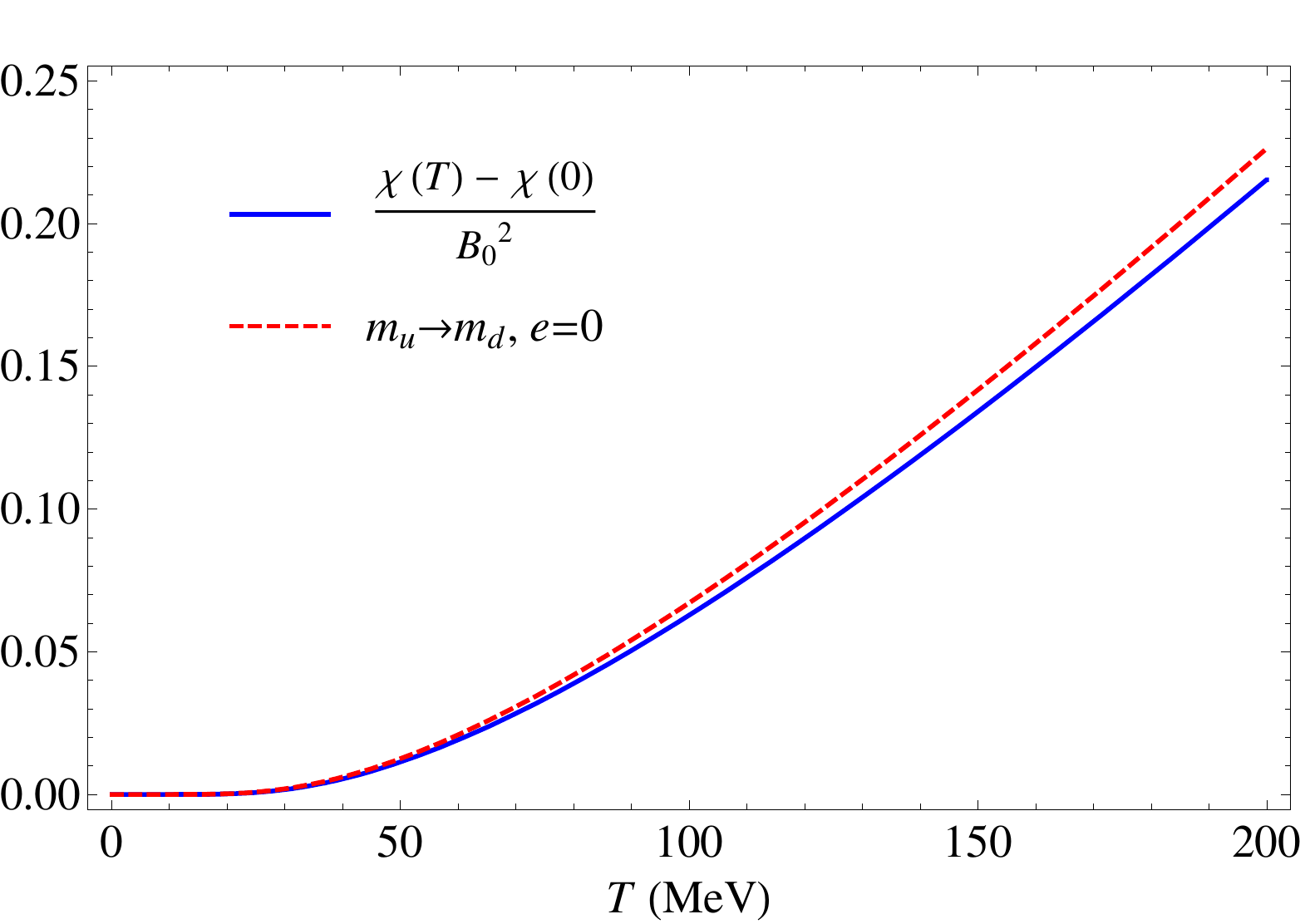}\hspace*{0.2cm} \includegraphics[scale=.52]{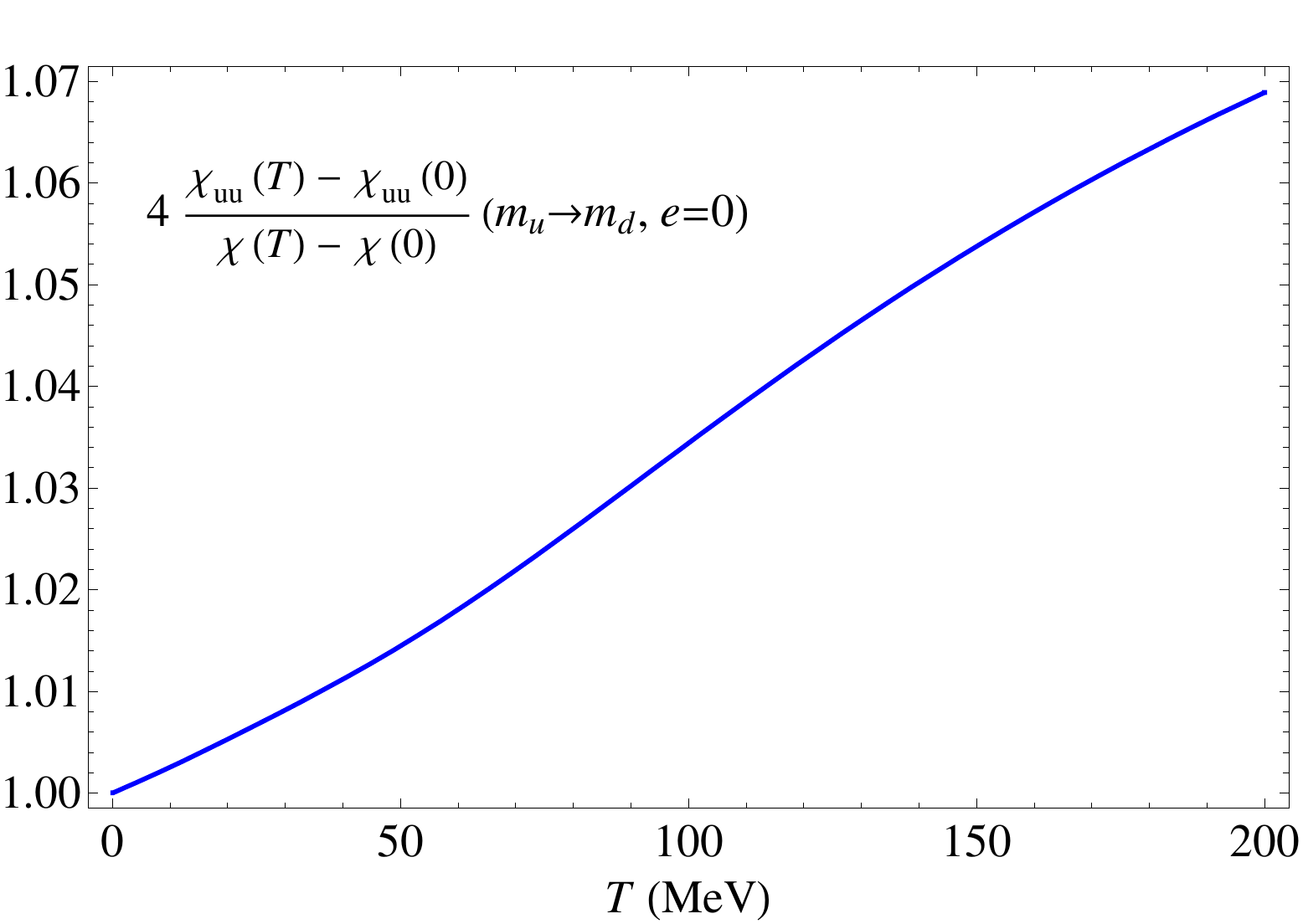}
 \caption{\rm \label{fig:susc} Temperature dependence of disconnected, connected and total susceptibilities in three-flavor ChPT, for physical quark and meson
  masses. The full blue curves show the results with all the isospin-breaking corrections of higher order included, while in the  dashed red ones we display the leading order in the isospin limit. We also show the IR expressions for the connected and disconnected parts, as well as the deviations of $4\chi_{uu}$ from $\chi_{tot}$ in the isospin limit (last plot). The $T=0$ results are given in Table \ref{tab:susczero}.}
\end{figure}

In Figure \ref{fig:susc} we plot our numerical ChPT results for the susceptibilities, including all the isospin-breaking corrections. The plots in that figure show the difference with respect to the $T=0$ results, which are collected in Table \ref{tab:susczero}. At $T=0$ we use the same LEC values as in \cite{Nicola:2010xt}, which are quoted in the table. Remember that the susceptibilities are independent of the EM LEC and that the disconnected one is independent of contact terms. The contact LEC $H_2^r$ appearing in the connected contribution is estimated from resonance saturation arguments \cite{Ecker:1988te,Amoros:2001cp}. The normalization used $B_0^2=M_{\pi}^4/(4\hat m^2)$  is the same  one used in some lattice works \cite{Aoki:2009sc}.

\begin{table}[h]
\begin{tabular}{|c|c|c|c|c|}
\hline
  & $\chi_{dis}/B_0^2$ & $\chi_{con}/B_0^2$ & $\chi/B_0^2$ & $4\chi_{uu}/B_0^2$   \\  \hline
 $m_s/\hat{m}=24$ &0.024  & 0.025 & 0.146 & 0.196 \\ \hline
$m_s/\hat{m}=24$, IL &0.025  & 0.025 & 0.148 & 0.197 \\ \hline
$m_s/\hat{m}=10$&0.016&0.024&0.113&0.163\\ \hline
$m_s/\hat{m}=10$, IL&0.017&0.023&0.114&0.161\\ \hline
$m_s/\hat{m}=100$&0.036&0.026&0.194&0.245\\ \hline
$m_s/\hat{m}=100$, IL&0.038&0.025&0.203&0.254\\ \hline
 \end{tabular}
 \caption{\rm \label{tab:susczero} $T=0$ values for the different susceptibilities in ChPT.  The isospin limit (IL) values correspond to $m_u\rightarrow m_d$ and $e=0$. For the third to sixth  rows we fix $m_u/m_d$ and $m_s$ and vary the light to heavy quark mass ratio. The first and second rows correspond to the physical values. The LEC values
 used are $H_2^r=2L_8^r=1.24\times 10^{-3}$, $L_6^r=0$ at the scale $\mu=770$ MeV.}
\end{table}

We see in the plots that the general features explored in our previous analytical discussion are well reproduced. First, the $\Od(e^2)$ and $\Od(m_d-m_u)^2$ terms neglected in (\ref{sustotsu3T})-(\ref{sustotsu3zero}) and (\ref{chidisTlo})-(\ref{chiconzerolo}) are numerically small for  the relevant temperature range, for physical values of quark and meson masses. In fact, as anticipated in our previous discussion, we see that those isospin corrections  are larger for the disconnected than the for connected part and are also larger with temperature, all due to the appearance of IR terms proportional to the second derivative of the tadpole in $\chi_{dis}$. Remember that the leading order in the isospin limit comes actually from the $\Od(m_d-m_u)$ terms in the condensates. Second, we appreciate qualitatively the linear and quadratic growth with temperature of the disconnected and connected parts respectively, as expected from the infrared analysis. In fact, we  see that although the connected term grows faster, its absolute value is much smaller due to the $M_\eta^2$ scale compared to the $M_\pi^2$ of the disconnected part.  However, it is important to note that the IR limit expressions (\ref{chidisTir})-(\ref{chiconTir}) are numerically rather far from the exact ones for the physical pion mass. The difference is larger for the disconnected contribution, since, as stated above, in (\ref{chidisTir}) we are neglecting $\Od(\log M_\pi)$ terms, while in (\ref{chiconTir}) the neglected terms are $\Od(M_\pi)$.  In fact, this justifies further our present  analysis, since we provide the full expressions  beyond the chiral limit. Also as discussed above, the infrared limit expressions for $T=0$ given in (\ref{chidiszeroir})-(\ref{chiconzeroir}) survive not only the chiral limit but also the $m_s\rightarrow\infty$ one, $\chi_{dis}$ still diverging but only logarithmically, which for physical masses makes the two susceptibilities numerically comparable. This is clearly seen in the values given in the first two rows of Table \ref{tab:susczero}. Actually, for this very same reason, and following our previous discussion, the deviations of $\chi_{uu}$ from the naive isospin-limit expectation $\chi/4$ is much more pronounced at $T=0$ than for finite $T$, as it can be seen by comparing the last two columns in Table \ref{tab:susczero} which give about a 30\%
 relative difference, while the last plot in Figure \ref{fig:susc} where we compare their thermal differences give only corrections below 10\%. In fact, this is consistent with our previous discussion about the influence of the connected part in the scaling properties observed in the lattice. If we consider the subtracted susceptibilities as defined in \cite{Ejiri:2009ac} from the subtracted condensate $\condu-(m_u/m_s)\conds$, we see that the dependence on the LEC disappears in the $\hat m/m_s\rightarrow 0$ limit. Remember that in this limit all the $T=0$ contribution of the connected part is absorbed in $h_3^r$ (see our previous comments) and therefore considering the subtracted susceptibility is equivalent to switch off the dominant $T=0$ part of the connected susceptibility. This is indeed observed in the lattice \cite{Ejiri:2009ac} since the subtracted susceptibility fits better the expected $O(N)$ scaling behaviour than the unsubtracted one.

\begin{figure}[h]
\includegraphics[scale=.53]{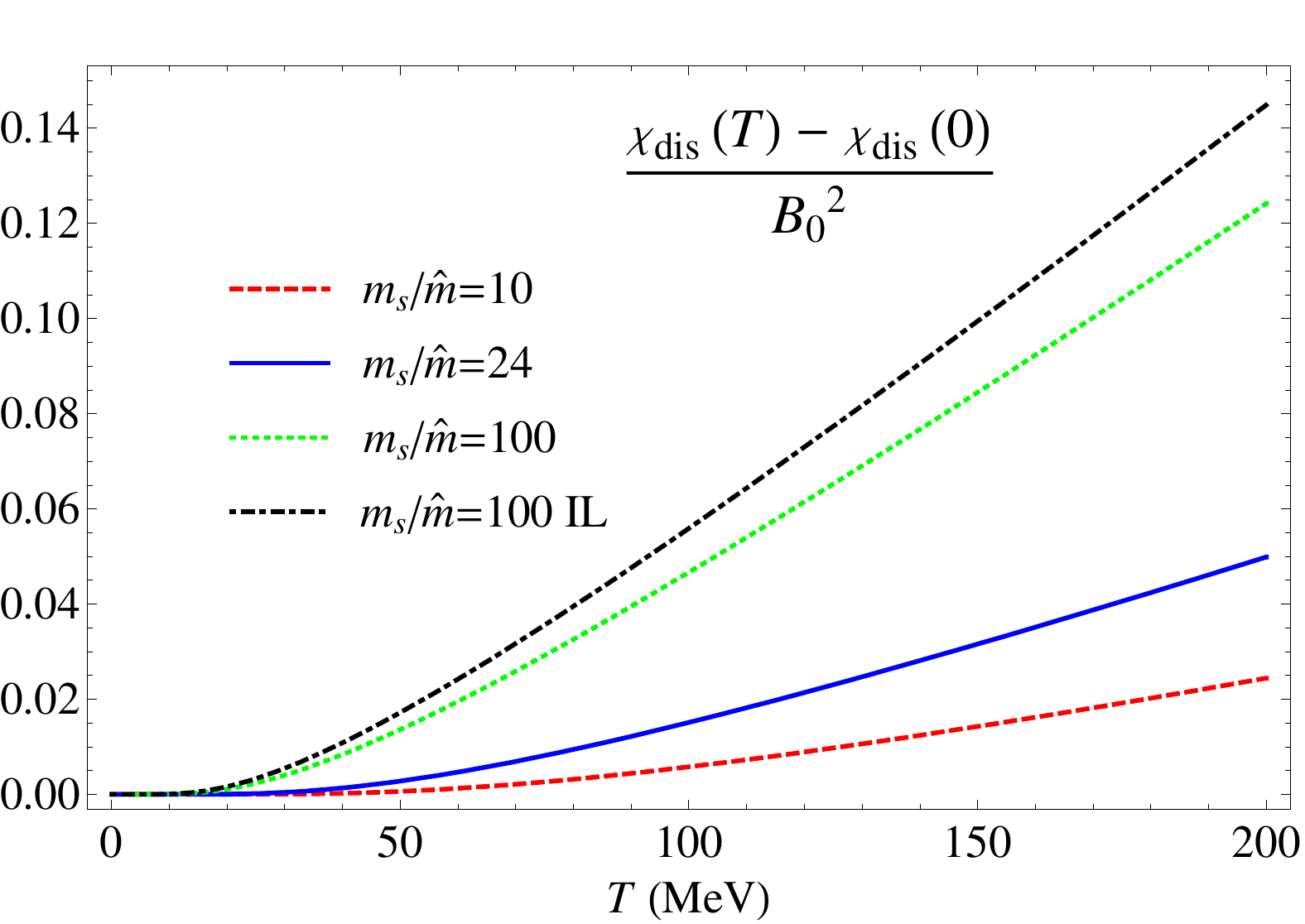}\hspace*{0.2cm}\includegraphics[scale=.53]{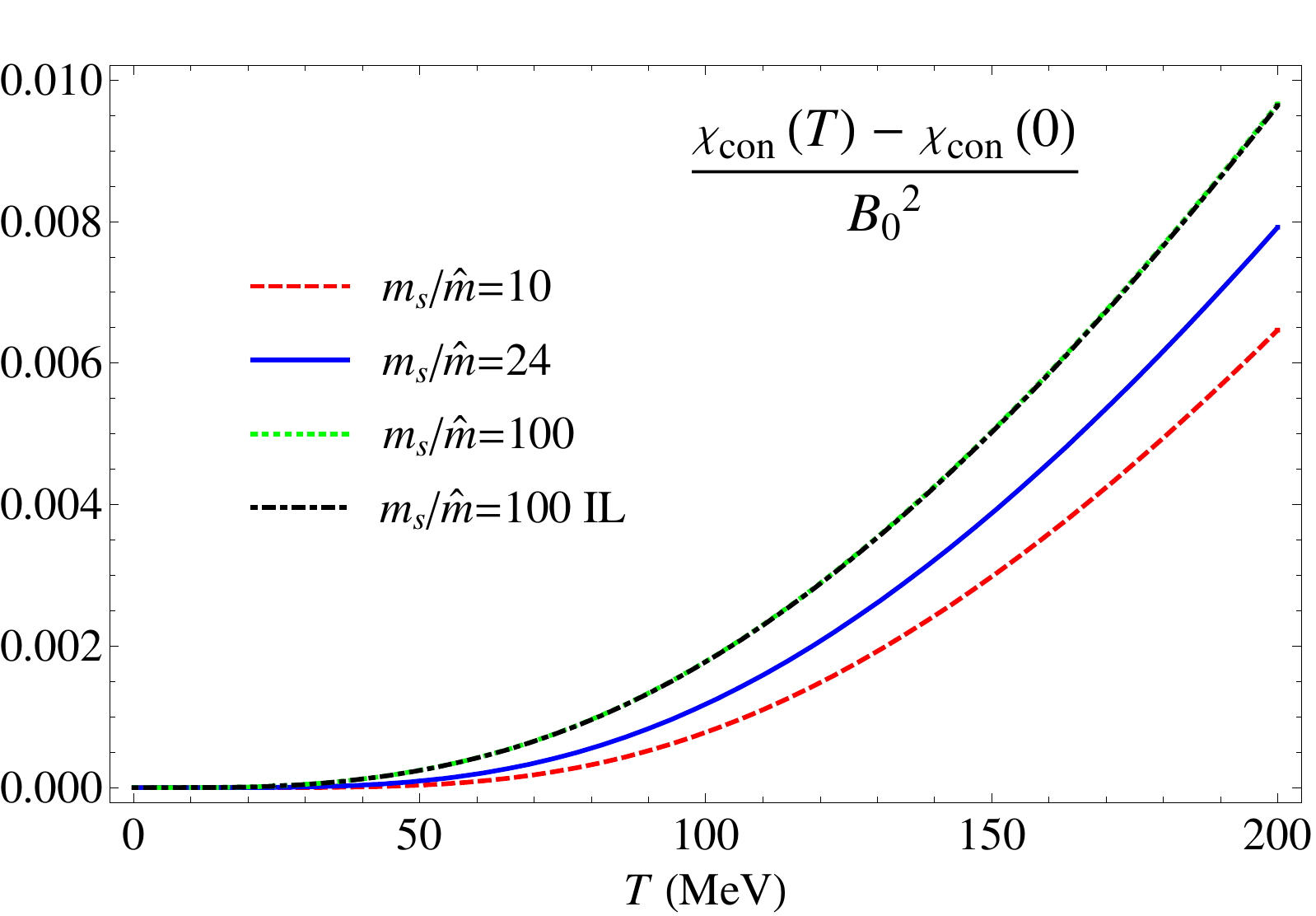}
\includegraphics[scale=.5]{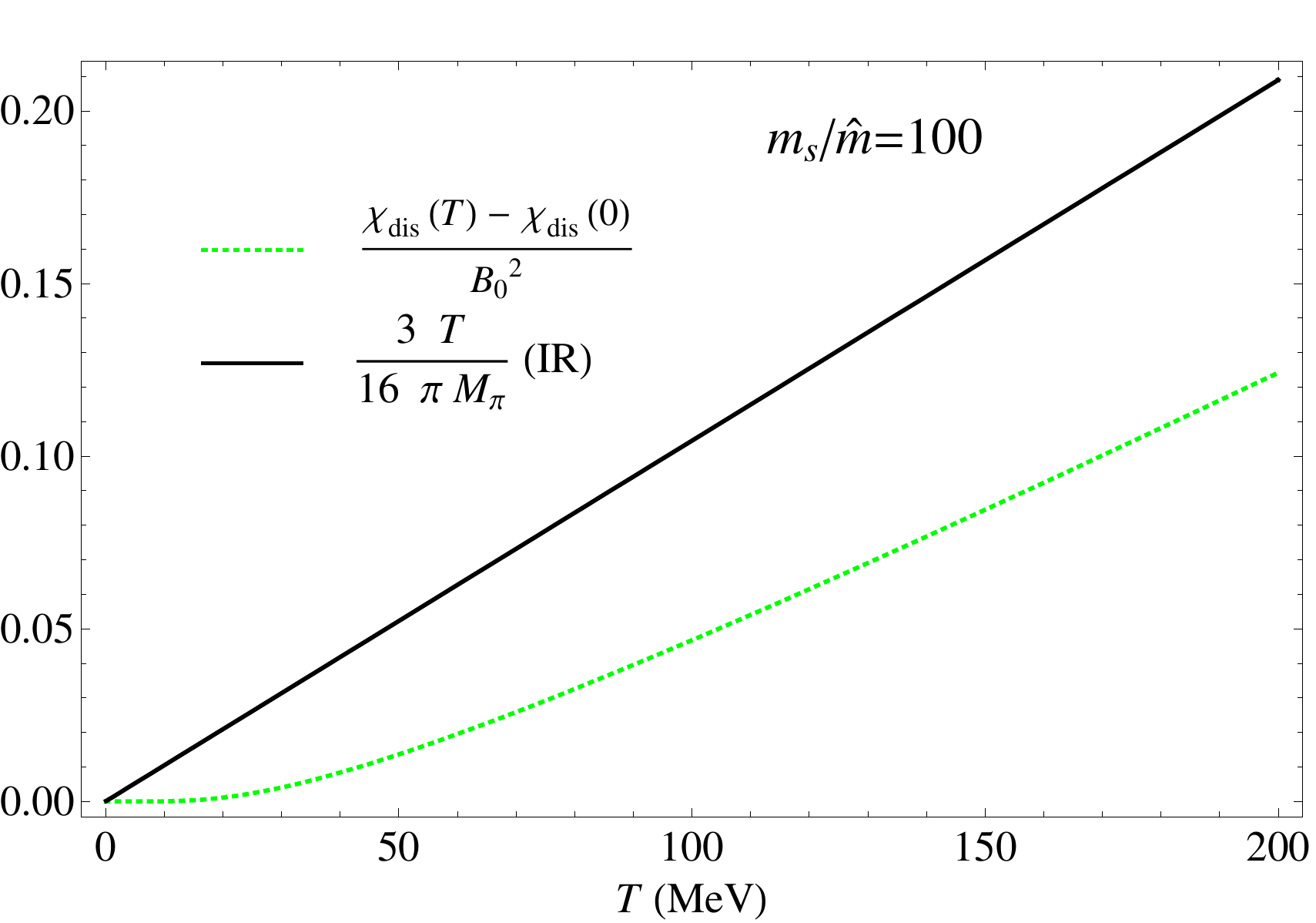}\hspace*{0.2cm}\includegraphics[scale=.5]{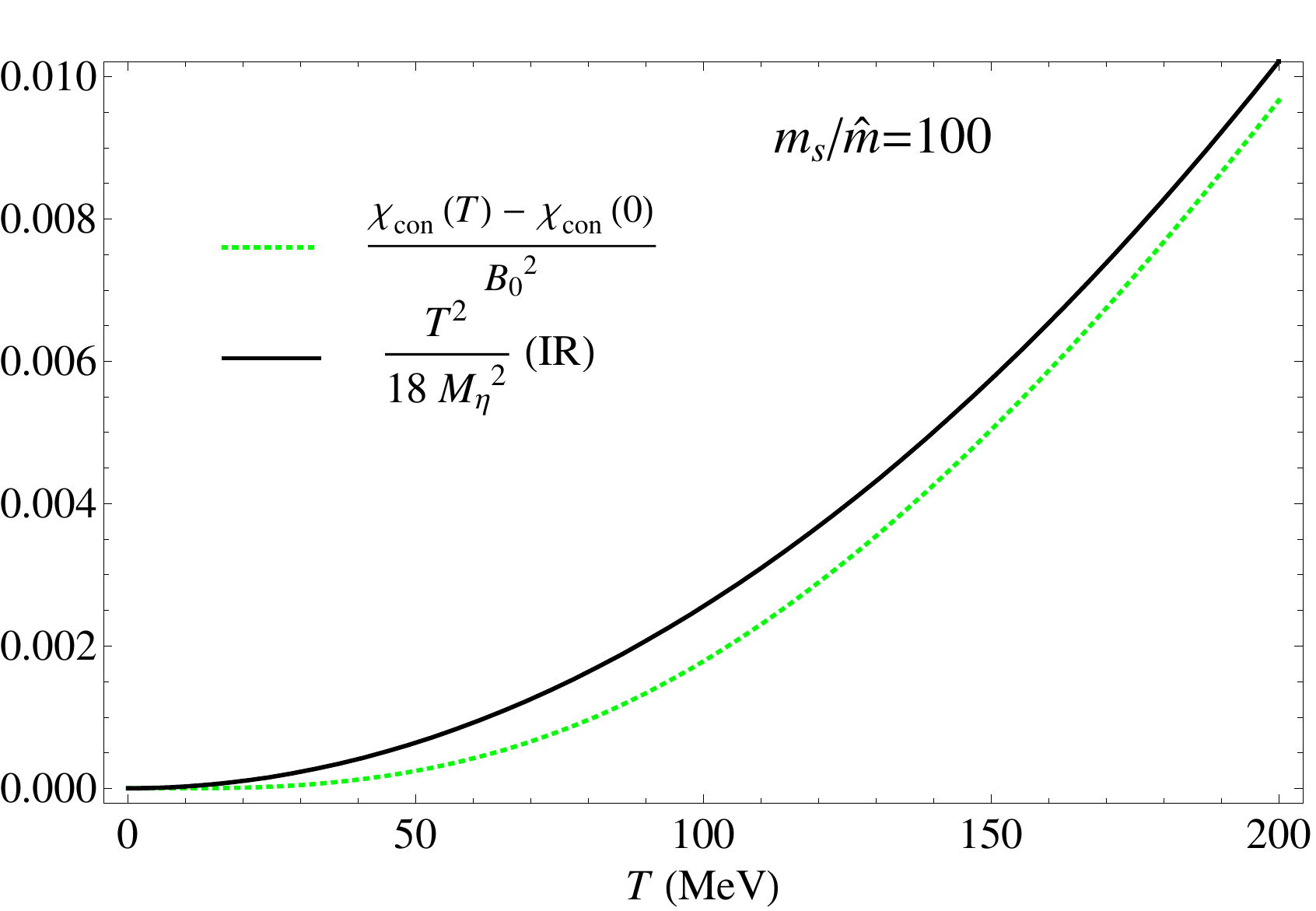}
 \caption{\rm \label{fig:suscmass} Quark mass dependence of the thermal disconnected, connected and total susceptibilities in three-flavor ChPT, for fixed $m_s$ and fixed $m_u/m_d$. We also show the isospin limit (IL) $m_u\rightarrow m_d$ and $e^2=0$ in the smaller light mass case. The $T=0$ results are given in Table \ref{tab:susczero}. For comparison, we also display the infrared limit in the $m_s/\hat m=100$ case.}
\end{figure}

The variation with the quark mass is displayed for $T\neq 0$ in Figure \ref{fig:suscmass} and for $T=0$ in Table \ref{tab:susczero}. It is important to remark that we have chosen to keep fixed the ratio $m_u/m_d\simeq 0.46$ (same value used in our previous analysis \cite{Nicola:2010xt}) and $m_s$, while we vary $\hat{m}/m_s$ above and below the physical quark mass ratio. In other words, when $\hat{m}\rightarrow 0^+$, $M_\pi\rightarrow 0^+$ while $M_{K,\eta}$ remain fixed. This is meant to be the relevant limit when approaching chiral restoration. In addition, since we can write $(m_d-m_u)/m_s=2(\hat{m}/m_s)(1-m_u/m_d)(1+m_u/m_d)^{-1}$, $\varepsilon$ scales in this limit as $\Od(\hat{m})$.  Recall that, although the values of the $L_i^r$ are fitted to low-energy data with physical masses \cite{Amoros:2001cp}, those LEC are formally independent of the quark masses \cite{Gasser:1984gg}. The same applies to the tree level value of $F$ we are using.

As we expected from our previous IR analysis in eqns. (\ref{chidisTir}) and (\ref{chiconTir}), the light quark mass dependence of the thermal disconnected susceptibility is much stronger than the connected one, as seen clearly in Fig.\ref{fig:suscmass} and as long as we take the limit in the order specified above. In terms of chiral restoration, this anticipates a much stronger growth or peak near $T_c$ for the disconnected part. From the same arguments. the behaviour of the connected part is expected to be softer near the transition, although growing with $T^2$ for low and moderate temperatures. We also show in the figure the comparison with the infrared limit for the smaller $\hat m/m_s$ case, where it can be seen that the curves are now closer than for the physical pion mass case in Figure \ref{fig:susc}.

In addition, the isospin corrections  are also more important for the disconnected part, where they actually increase as $\hat{m}$ is decreased, than for the connected one, where the isospin limit and complete curves are almost indistinguishable  in the figure. The same holds for the $T=0$ contributions in Table \ref{tab:susczero}. According to our previous discussion, this behaviour of the isospin corrections arises from the dominant IR terms $\chi_{dis}\sim B_0^2(TM_\eta^2/M_\pi^3)\varepsilon^2=\Od(\sqrt{\hat m})$ as compared to $\chi_{con}\sim B_0^2(T/M_\pi)\varepsilon^2=\Od(\hat m^{3/2})$. This effect is weaker for the $T=0$ contributions since the IR leading corrections $\nu_\pi\varepsilon^2$ (in $\chi_{con}$) and $(\partial \nu_\pi/\partial M_\pi^2)\varepsilon^2$ (in $\chi_{dis}$) diverge softly, namely, as $\hat m^2\log \hat m$ and $\hat m$ respectively. Thus, although the isospin corrections are amplified in the disconnected susceptibility for large temperatures and small masses, they are still perturbatively under control in the chiral limit.

The limit where $\hat m/m_s$ vanishes not by taking $\hat m\rightarrow 0^+$ but keeping $\hat m$ fixed and taking $m_s\rightarrow\infty$ is where we recover the pure $SU(2)$ results, as discussed before. In that case, it is the connected part which is more sensitive to the quark mass variation, vanishing for large $M_\eta^2$, while the disconnected one remains invariant. Although this is formally interesting for connecting the $SU(2)$ and $SU(3)$ cases, it is not so relevant for studying the critical behaviour.

\section{Conclusions}

In this work we have analyzed the relevant observables regarding chiral symmetry restoration, namely quark condensates and scalar susceptibilities, in the presence of isospin breaking. We have considered on the same footing the QCD ($m_u/m_d$ mass difference) and electromagnetic corrections, to one loop in Chiral Perturbation Theory, both in the $SU(2)$ and $SU(3)$ sectors. Our analysis provides useful and model-independent results regarding several  relevant aspects of isospin breaking and chiral restoration, which may be particularly  interesting  for lattice studies.

The sum $\condsum_T$, the order parameter for chiral restoration, receives small isospin-breaking corrections for the physical values of masses and electric charge. These corrections affect only slightly, less than $1\%$, to the value of the critical temperature, which they increase as a ferromagnetic response. The difference $\conddif_T$ is the order parameter of isospin breaking. It is temperature independent in the $SU(2)$ limit, but when kaons and eta are included, it shows an increasing behaviour, which in the chiral limit is given by  $(m_u-m_d)T^2/M_\eta^2$. The deviations with respect to its $T=0$ value become sizable as the temperature is increased, but they are controlled by a larger energy scale $M_\eta^2$ than the typical $F_\pi^2$ of $\condsum_T$. This large growth of isospin breaking does not reflect in the  chiral restoration temperatures of $\condu_T$ and $\condd_T$, which  remain  close to each other, consistently with the idea that chiral restoration is little affected. We have also evaluated the temperature corrections to the sum rule relating the $\conds_T/\condu_T$ and $\condu_T/\condd_T$ ratios, which is useful because it does not involve undetermined contact low-energy constants. The corrections in this case come directly from the $\condd_T/\condu_T$ ratio and are therefore rather large for the temperatures of interest.

A very important part of the present work has been the analysis of scalar susceptibilities in the isospin asymmetric scenario. We have related the different flavor susceptibilities with the total, quark connected and quark disconnected susceptibilities often used in lattice analysis. Electromagnetic corrections to the quark condensate turn out to be directly related by a sum rule to the total susceptibility and then to the growth of fluctuations, which is meant to be maximum near the critical point. This sum rule is valid for any small deviation of the pion masses, as for instance the one arising in the staggered lattice formalism due to  taste breaking effects. Actually, we have made rough estimates of the corrections to condensates expected from this source, comparing lattices of different sizes among them and with the continuum limit. These estimates are in good agreement with the errors quoted in the lattice works.

The isospin asymmetric calculation  allows for a direct extraction of the  connected and disconnected susceptibilities, even in the isospin symmetric limit. The terms in $\conddif_T$ linearly proportional to $m_u-m_d$ give  contributions to the connected part not vanishing in the isospin limit and which affect for instance the naive extrapolation of a given flavor susceptibility to the total one. Our analysis  provides  model-independent predictions for the mass, temperature and isospin dependence of those quantities, which should be recovered in lattice analysis as they approach the continuum limit. In accordance with the behaviour of the corresponding order parameters, the disconnected susceptibility  shows a linear growth  at low and moderate temperatures, infrared divergent near the chiral limit as $T/M_\pi$, whereas the connected one  is infrared regular but survives the chiral limit as a growing $T^2/M_\eta^2$ behaviour. The chiral or infrared limit gives qualitatively the behaviour as the temperature approaches chiral restoration but numerically is not a good approximation for physical pion masses. The higher order isospin breaking corrections are quadratic in $m_u-m_d$ and are enhanced in the chiral limit for the disconnected susceptibility, as long as $m_s$ and $m_u/m_d$ remain fixed. The ChPT susceptibilities reproduce the growing $T$-dependence at low and moderate temperatures in a model-independent way. Although they do not show the peaks  expected near the transition, our small mass analysis allows to infer that the disconnected part should have a more pronounced peak than the connected one, the latter expected to present a rather soft behaviour. This difference  can be interpreted from the different order parameters that fluctuate in each case: the chiral quark condensate for the disconnected piece and the isospin-breaking one in the connected case. In the formal $SU(2)$ limit $m_s\rightarrow\infty$  the connected contribution becomes  temperature independent, like $\conddif_T$.  Our analysis for the susceptibilities is consistent with previous related work in the literature.

\section*{Acknowledgments}
We are grateful to W.Unger  for useful comments. Work partially supported by the Spanish
research contracts FPA2008-00592,  FIS2008-01323, UCM-BSCH GR58/08 910309 and the FPI programme (BES-2009-013672).

\end{document}